\documentclass{article}

\usepackage[onecolumn]{emulateapj}

\newcommand{\rstar}{r^{*}}

\newcommand{\rf}[1]{\ref{fig:#1}}
\newcommand{\secc}[1]{\S \ref{sec:#1}}
\newcommand{\secl}[1]{\label{sec:#1}}

\newcommand{\tskip}{}

\newcommand{\be}{\begin{equation}}
\newcommand{\ee}{\end{equation}}

\newcommand{\ec}[1]{Eq.~[\ref{eq:#1}]}
\newcommand{\eql}[1]{\label{eq:#1}}

\def\matk{{\bf K}}
\def\wind{{\cal W}}

\input epsf
\begin{document}
\title{The 3D Power Spectrum from Angular Clustering of Galaxies in
Early SDSS Data\altaffilmark{1}}

\author{ Scott Dodelson\altaffilmark{2,3},
Vijay K. Narayanan\altaffilmark{4},
Max Tegmark\altaffilmark{5},
Ryan Scranton\altaffilmark{2,3},
Tamas Budavari\altaffilmark{6},
Andrew Connolly\altaffilmark{7},
Istvan Csabai\altaffilmark{6},
Daniel Eisenstein\altaffilmark{8},
Joshua A. Frieman\altaffilmark{2,3},
James E. Gunn\altaffilmark{4},
Lam Hui\altaffilmark{9},
Bhuvnesh Jain\altaffilmark{5},
David Johnston\altaffilmark{2,3},
Stephen Kent\altaffilmark{2,3},
Jon Loveday\altaffilmark{10},
Robert C. Nichol\altaffilmark{11}, Liam O'Connell\altaffilmark{10},
Roman Scoccimarro\altaffilmark{12,13},
Ravi K. Sheth\altaffilmark{2}, Albert Stebbins\altaffilmark{2},
Michael A. Strauss\altaffilmark{4}, Alexander S. Szalay\altaffilmark{6},
Istv\'an Szapudi\altaffilmark{14},
Michael S. Vogeley\altaffilmark{15},
Idit Zehavi\altaffilmark{2},
James Annis\altaffilmark{2},
Neta A. Bahcall\altaffilmark{4},
Jon Brinkman\altaffilmark{16},
Mamoru Doi\altaffilmark{17},
Masataka Fukugita\altaffilmark{18},
Greg Hennessy\altaffilmark{19},
\v{Z}eljko Ivezi\'c\altaffilmark{4},
Gillian R. Knapp\altaffilmark{4},
Peter Kunszt\altaffilmark{20},
Don Q. Lamb\altaffilmark{3},
Brian C. Lee\altaffilmark{2},
Robert H. Lupton\altaffilmark{4}, Jeffrey A. Munn\altaffilmark{19},
John Peoples\altaffilmark{2},
Jeffrey R. Pier\altaffilmark{19}, Constance Rockosi\altaffilmark{3},
David Schlegel\altaffilmark{4}, Christopher Stoughton\altaffilmark{2},
Douglas L. Tucker\altaffilmark{2},
Brian Yanny\altaffilmark{2}, Donald G. York\altaffilmark{3,21}
for the SDSS Collaboration}

\altaffiltext{1}{Based on observations obtained with the Sloan Digital Sky
Survey}
\altaffiltext{2}{Fermi National Accelerator Laboratory, P.O. Box 500, Batavia,
IL 60510, USA}
\altaffiltext{3}{Astronomy and Astrophysics Department, University of
Chicago, Chicago, IL 60637, USA}
\altaffiltext{4}{Princeton University Observatory, Princeton, NJ 08544,
USA}
\altaffiltext{5}{Department of Physics, University of Pennsylvania,
Philadelphia, PA 19104, USA}
\altaffiltext{6}{Department of Physics and Astronomy, The Johns Hopkins
University, 3701 San Martin Drive, Baltimore, MD 21218, USA}
\altaffiltext{7}{University of Pittsburgh, Department of Physics and
Astronomy, 3941 O'Hara Street, Pittsburgh, PA 15260, USA}
\altaffiltext{8} {University of Arizona}
\altaffiltext{9}{Department of Physics, Columbia University, New York, NY
10027, USA}
\altaffiltext{10}{Sussex Astronomy Centre, University of Sussex, Falmer,
Brighton BN1 9QJ, UK}
\altaffiltext{11}{Department of Physics, 5000 Forbes Avenue, Carnegie
Mellon
University, Pittsburgh, PA 15213, USA}
\altaffiltext{12}{Department of Physics, New York University, 4 Washington
Place, New York, NY 10003}
\altaffiltext{13}{Institute for Advanced Study, School of Natural
Sciences,
Olden Lane, Princeton, NJ 08540, USA}
\altaffiltext{14}{Institute for Astronomy, University of Hawaii, 2680
Woodlawn Drive, Honolulu, HI 96822, USA}
\altaffiltext{15}{Department of Physics, Drexel University, Philadelphia,
PA
19104, USA}
\altaffiltext{16}{Apache Point Observatory, P.O. Box 59, Sunspot, NM 88349-0059}
\altaffiltext{17}{Dept. of Astronomy and Research Center for
the Early Universe, School of Science, University of Tokyo, Tokyo
113-0033, Japan}
\altaffiltext{18}{Institute for Cosmic Ray Research, University of
Tokyo, Kashiwa 277-8582, Japan}
\altaffiltext{19}{U.S. Naval Observatory, Flagstaff Station, P.O. Box 1149, Flagstaff,
      AZ 86002-1149}
\altaffiltext{20}{CERN, IT Division, Database Group, 1211 Geneva 23,
 Switzerland }
\altaffiltext{21}{Enrico Fermi 
Institute, 5640 So. Ellis Ave., Chicago, IL 60637}

\begin{abstract}
Early photometric data from the Sloan Digital Sky Survey (SDSS)
contain angular positions for 1.5 million galaxies. 
In companion papers, the angular correlation function $w(\theta)$ 
and 2D power spectrum $C_l$ of these galaxies are presented. 
Here we invert Limber's equation to extract the 3D power spectrum from 
the angular results. 
We accomplish this using an estimate of $dn/dz$, the redshift distribution 
of galaxies in four different magnitude slices in the SDSS photometric catalog.
The resulting 3D power spectrum estimates from $w(\theta)$ and $C_l$ agree 
with each other and with previous estimates over a range in wavenumbers
$0.03 < k/{\rm h\ Mpc}^{-1} < 1$. 
The galaxies in the faintest magnitude bin ($21 < \rstar < 22$, which have 
median redshift $z_m=0.43$) are less clustered than the galaxies in the
brightest magnitude bin ($18 < \rstar < 19$ with $z_m=0.17$), especially on 
scales where nonlinearities are important. 
The derived power spectrum agrees with that of Szalay et al. (2001) who 
go directly from the raw data to a parametric estimate of the power spectrum. 
The strongest constraints on the shape parameter
$\Gamma$ come from the faintest galaxies (in the magnitude bin 
$21 < \rstar < 22$), from which we infer 
$\Gamma = 0.14^{+0.11}_{-0.06}$ ($95\%$ C.L.). 

\end{abstract}

\section{Introduction}

The statistical properties of galaxy clustering in the Universe encode
a wide variety of cosmological information. The most powerful example of this
is the 3D power spectrum, which theoretically depends on a number of 
cosmological parameters, notably the matter density, the Hubble constant, 
the baryon density, and the density and mass of neutrinos.
Observationally, the power spectrum can be probed in a number of ways. While
a spectroscopic survey  contains information about the radial positions of
galaxies, this information is (a) difficult to obtain for a large sample
of galaxies over a cosmological volume, and (b)  difficult to
interpret because of peculiar velocities. 
Angular (imaging) surveys do not suffer from these problems, but of course are 
fundamentally limited to probing a projection of the 3D galaxy distribution
on the sky. 
Often the vast numbers of galaxies in 2D surveys enable them
to probe the 3D power spectrum more effectively than redshift surveys.

These generalities are illustrated by the Sloan Digital Sky Survey 
(SDSS) (York et al. 2000).
The early data release (Stoughton et al. 2001) contains of order $50,000$ 
redshifts in the spectroscopic survey, but several million galaxies in 
the photometric survey. Companion papers (Scranton et al. 2001, 
hereafter Sc01; Connolly et al. 2001, C01; Tegmark et al. 2001, T01) use these
results to measure the angular correlations in the 
early data; here we extract the 3D power spectrum.

To estimate the 3D power spectrum, we need to understand how the
3D galaxy distribution is projected onto the 2D sky (Limber 1953). 
The most important ingredient in this projection is the redshift distribution 
$dn/dz$ of galaxies. 
Given the redshift distribution appropriate to a given survey and magnitude
limit, one can form the kernels for
either the correlation function $w(\theta)$ or the 2D power
spectrum, the $C_l$'s. The kernels convolved with the 3D power
spectrum yield the angular correlations. Extracting the 3D
power spectrum therefore entails
inverting the kernel. While the $C_l$'s are simpler to invert
due to a more straightforward kernel and a covariance matrix that is
more nearly diagonal, $w(\theta)$ has been more traditionally used. 
Besides the advantage of tradition, working in real space has enabled us 
to better understand possible systematic effects in 
the galaxy sample (see Sc01) and allowed us to probe the correlations
on smaller scales.
So here we extract the 3D power spectrum from both $w(\theta)$ and $C_l$,
and check the consistency of the two estimates.

As described in Sc01, C01, and T01, the galaxies in the early SDSS data
are divided into four magnitude bins, in the range $18 < \rstar < 22$.
\footnote{Following Stoughton et al. 2001 , we refer to the SDSS passbands as
$u$, $g$, $r$, $i$ and $z$. As the SDSS photometric calibration
system is still being finalized, the SDSS photometry presented here
is referred to as $u^*$, $g^*$, $r^*$, $i^*$ and $z^*$.}
The redshift distribution of galaxies in each of these four magnitude
bins is derived in \secc{dndz} using three sets of observations:
the luminosity function inferred from the Canadian
Network for Observational Cosmology Field Galaxy Redshift Survey (Lin et al.
1999; hereafter CNOC2); a set of galaxies with redshifts from the Canada 
France Redshift Survey (CFRS) that were also observed photometrically in SDSS; 
and photometric redshifts from the SDSS photometric data themselves. 
The resultant kernels are discussed in \secc{kernels}. 
Inversion to extract the 3D power spectrum is presented in \secc{invert}. 

This paper is one of a series of papers analyzing angular clustering in 
early SDSS data. The basis for all of these results is the systematics 
paper of Sc01. The relationship between the papers in this series (T01;
C01; Szalay et al. 2001, hereafter Sz01) is depicted schematically in Figure \rf{relation}.
Sz01 goes directly from the data to estimate the 
parameters of the 3D power spectrum using a likelihood approach. This
technique has the advantage of not depending on the covariance matrix of
$w(\theta)$ (or $C_l$) which is difficult to compute due to non-Gaussianities. 
It assumes that galaxy overdensities are Gaussian distributed, which 
is accurate only on very large scales.
A related point is that they do not work directly in Fourier space where 
the division between linear and non-linear scales is transparent. 
The two-step method of the other papers (data to two-point functions to 
$P(k)$ parameters) should, in principle, yield the same result as the 
likelihood approach. Its strengths and weaknesses are complementary to
the likelihood approach. In \secc{conclusions} of this paper, we compare 
$P(k)$ derived using these very different techniques.

\begin{figure}[t]
\centerline{\epsfxsize=3.5truein\epsffile{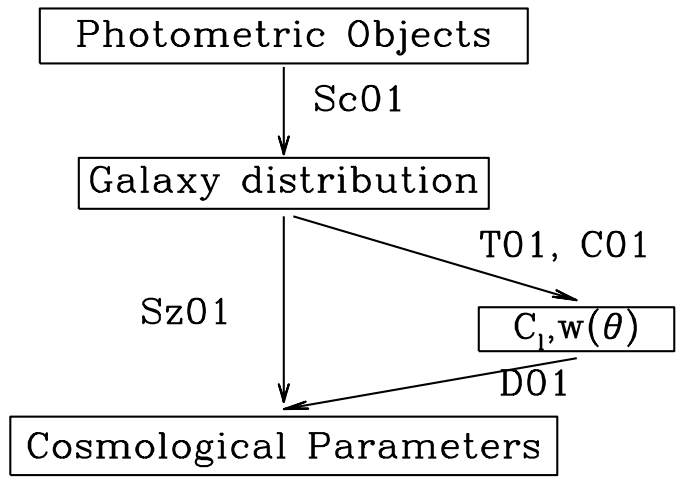}}
\caption{The papers analyzing early SDSS angular clustering. All work is
based on the systematics paper of Scranton et al. (Sc01), which goes from
photometric objects and positions to a map with galaxy overdensities 
and noise, and shows that
these data are free from
systematic effects for galaxies brighter than $\rstar= 22$.
Tegmark et al. (T01) and Connolly et al. (C01) estimate two-point functions, 
$C_l$ and $w(\theta)$ respectively, from the galaxy angular positions. 
Szalay et al. (Sz01) go directly from the data to estimating the parameters
characterizing the 3D power spectrum.
The current paper (D01) builds on the estimated angular two-point functions 
to infer the 3D power spectrum, and compares it with the direct 
data-to-parameter approach of Sz01.}
\label{fig:relation}
\end{figure}

\section{Redshift Distribution}
\label{sec:dndz}

The projection of the 3D power spectrum onto the plane of the sky depends on
the radial selection function. If an imaging survey is very deep, then 
a fixed angular scale corresponds to very large physical scales.
Determining the redshift distribution of galaxies in different magnitude 
bins is therefore a crucial step towards interpreting the angular 
correlations. SDSS is omewhat limited in this regard, as (at least for
normal galaxies) it only takes redshifts of galaxies brighter than $\rstar\sim19$.
Therefore, we need to turn to external observations to estimate the radial
selection function.

Here we estimate $dn/dz$ using the three samples mentioned above 
(CNOC2, SDSS+CFRS, and SDSS photo-$z$). The redshift distribution from each 
sample is fit to the simple parametric form
(Baugh \& Efstathiou, 1993)
\be
{dn\over dz} = {3 z^2\over 2 (z_m/1.412)^3}\ \exp\left(-(1.412 z/z_m)^{3/2}\right)
.\eql{dndzpar}\ee
where the median redshift of the galaxy distribution is $z_m$,
and the integral of this $dn/dz$ over all redshifts is unity.
The early SDSS angular data we study here are divided into unit magnitude 
bins, the brightest including galaxies with $model$ magnitudes between 
$\rstar=18$ and $\rstar=19$ progressing to the  faintest with $21 < \rstar < 22$. 
(see Stoughton et al. 2001 for a description of the different magnitudes
measured by SDSS).
For each of these magnitude bins we determine the median redshift $z_m$. 
We will see that the $z_m$ measured from the three samples agree very well
 in all the four magnitude bins. 
To get an error on $z_m$, which will ultimately be propagated to an 
error on  cosmological parameters, we use the reported uncertainties in the 
CNOC2 sample. We argue that this error is very conservative.

\subsection{CNOC2}

The CNOC2 Survey (Lin et al. 1999) measured redshifts for over $2000$ galaxies 
in the redshift range $0.12 < z < 0.55$, distributed over $2490$ arcmin$^{2}$
of sky, in the magnitude range $17 < R_c < 21.5$.
They fit the galaxy luminosity function to a Schechter function
\be
\phi(M) = 0.4\ln(10) \phi^*(z=0) 10^{0.4 Pz} \left[10^{0.4(M^*-M)}\right]^{1+\alpha}
\exp\left\{ -10^{0.4(M^*-M)} \right\}
\eql{schechter}
\ee
where $\phi(M)$ is the differential luminosity function, a function of
absolute magnitude $M$ (in the rest frame $R_c$ band);
$\phi^*$ is the normalization; $M^*=M^{*(0)} - Qz$ is the characteristic
magnitude; $Q$ parameterizes the luminosity evolution of the galaxy population
with redshift,  and $\alpha$ is the faint end slope of the Schechter function.
CNOC2 fit luminosity functions separately to three different types of galaxies 
-- early, intermediate, and late;
for each galaxy type, they determine the best fit values of the Schechter 
parameters $M^{*(0)},Q,P,\alpha,$ and $\phi^*$. We use the central values 
for these parameters reported in Table 2 of CNOC2.

For each SDSS magnitude bin, we translate the CNOC2 Schechter
functions into $dn/dz$. This translation would be straightforward if all 
galaxies had flat spectra.
We could simply associate with each apparent magnitude bin a corresponding 
absolute magnitude bin, using the relation
\be
M = \rstar - 5 \log_{10}[\chi(z)(1+z)/10 {\rm pc}]
\eql{abssim}\ee
where $\chi(z)$ is the comoving distance out to redshift $z$.
Thus, a range of  apparent magnitudes translates into a different range of
absolute magnitudes at different redshifts. 
For example, in this unrealistic example of a flat galaxy spectrum, 
the $18 < \rstar < 19$ bin at redshift $z=0.2$ corresponds 
to $-21 < M < -20$ in a flat, matter-dominated Universe with Hubble
constant $H_0=100$km sec$^{-1}$ Mpc$^{-1}$.

Converting apparent magnitude in the $\rstar$ band to absolute magnitude in the 
$R_c$ band -- the band in the CNOC2 fit closest to $\rstar$ -- requires two 
modifications of \ec{abssim}. 
We need to convert the galaxy magnitude in $R_c$ to $\rstar$ and
we need to apply the $K-$ correction which accounts for the shift in the 
spectrum due to redshift (the $\rstar$ band of a high $z$ galaxy is sampling 
the spectrum at shorter restframe wavelengths than that of a $z=0$ galaxy).
These corrections change \ec{abssim} to
\be
R_c = \Delta + \rstar - 5 \log_{10}[\chi(z) (1+z)/10 {\rm pc}] - K(z)
\ee
where $\Delta$, the color difference between $\rstar$ and $R_c$, and
$K(z)$, the K-correction, both depend on the spectral type of the galaxy.
We adopt the values of $\Delta$ and $K(z)$ (in the SDSS $\rstar$ band) given in
Table 3, and Figure 20 of Fukugita et al. (1995), respectively.
Using this conversion to fix $R_{c,{\rm min}}$ and $R_{c,{\rm max}}$ in 
terms of the upper and lower apparent magnitude limits, 
we integrate the Schechter function in \ec{schechter} so that 
\be
{dn\over dz} \propto {\chi^2 d\chi/dz\over (1+z)^3} 
\int_{R_{c,{\rm min}}}^{R_{c,{\rm max}}}
dM \phi(M)
.\ee
The proportionality constant here is irrelevant, since the correlations 
do not depend on it. We choose it so that the integral of $dn/dz$ over all 
redshifts is unity.

Both the color difference $\Delta$ and the $K-$ correction
depend on galaxy type, as does the Schechter fit in CNOC2 itself. 
In a given apparent magnitude bin,
we summed the contribution from the three different galaxy ``types'': Early,
Intermediate, and Late, using the Schechter fit for that galaxy type
given in CNOC2.
For  $\Delta$ and $K(z)$, we associate E with Early, Sbc with
Intermediate, and Im with Late type galaxies. The individual contributions
from these three type are shown in Figure \rf{dndz_1920} for galaxies  in
the magnitude bin $19 < \rstar < 20$ (the same trend holds for all other bins).
Note that late-type galaxies are at lower redshift. This corresponds
to the realization by CNOC2 that the luminosity function of the
late-type galaxies has a steep faint end slope $(\alpha = -1.2)$,
and hence there are many low-luminosity late-type galaxies, which are
seen only at low redshifts.

\begin{figure}[thbp]
\centerline{\epsfxsize=5truein\epsffile{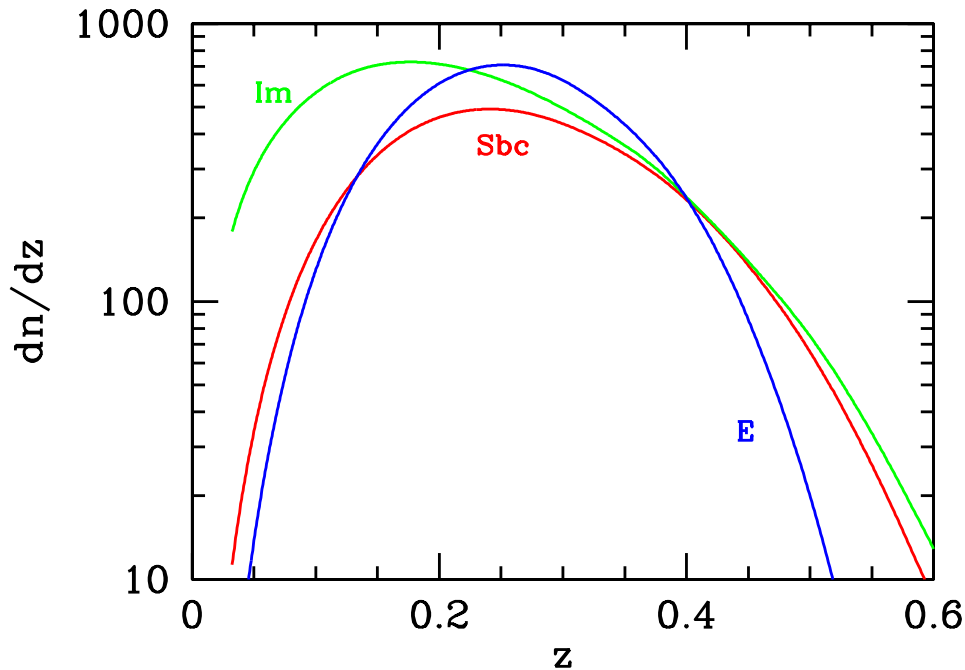}}
\caption{Contributions to the full $dn/dz$ from different galaxy types
in the apparent magnitude bin $19 < \rstar < 20$ 
(other bins show similar trends).
Each type is representative of one of the classes of galaxies for which
CNOC2 derived Schechter fits (Im with Late; Sbc with Intermediate; and E with
Early). Here the relative normalizations are important since the total $dn/dz$
(the blue curve in Figure \rf{dndz}) is the sum of these three individual 
contributions.}
\label{fig:dndz_1920}
\end{figure}

The total redshift distribution in a given magnitude bin is the sum
of the contributions from all three galaxy classes. The resulting
distributions are shown in Figure \rf{dndz} 
for the four magnitude bins under consideration. We will refer to
these as the {\it direct} CNOC2 $dn/dz$'s to distinguish them from the
fits to \ec{dndzpar}. As we will see shortly, \ec{dndzpar} is not a perfect
fit to the direct $dn/dz$.
However, we will show below that this difference between the direct 
$dn/dz$ and the empirical fit has a negligible effect on the derived 
power spectrum.

\begin{figure}[t]
\centerline{\epsfxsize=5truein\epsffile{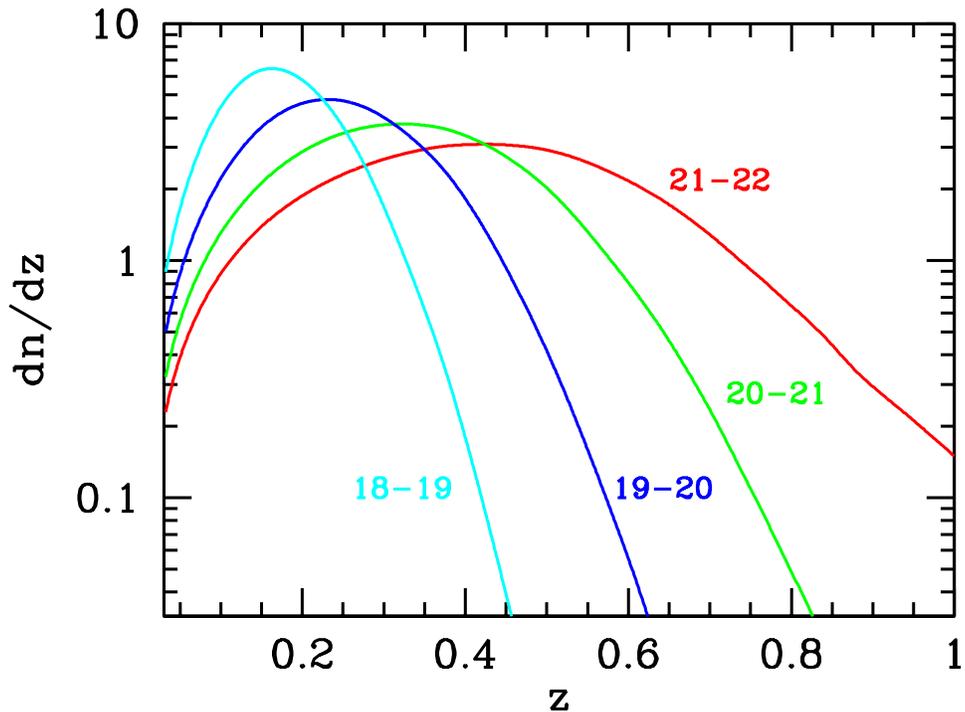}}
\caption{Galaxy redshift distributions in four apparent magnitude bins 
using the CNOC2 selection function. All curves here (unlike those
in Figure~\rf{dndz_1920}) are normalized so that 
$\int dz dn/dz=1$.}
\label{fig:dndz}
\end{figure}

Figures \rf{dndz_1920} and \rf{dndz} use the best-fit values of the
Schechter function parameters from CNOC2. These parameters undoubtedly 
have correlated  errors, although the correlation coefficients 
are not reported by CNOC2.
Therefore, we neglect the correlations between the Schechter parameters,
and derive a conservative estimate of the uncertainty in 
$dn/dz$ due to the uncertainties in the luminosity function, using one
hundred Monte-Carlo runs.
In any Monte-Carlo run, we choose each Schechter parameter from
a Gaussian distribution with mean equal to the best fit value and 
standard deviation equal to the error reported by CNOC2. In a given
run, we then compute the median redshift $z_m$ of the galaxy redshift 
distribution.
The second column in Table 1 shows the median redshift $z_m$ (obtained
from the best fit values of the Schechter parameters) and a standard deviation
for each magnitude bin;
the standard deviations are measured from the 
hundred Monte-Carlo runs. We emphasize that this is a very conservative 
estimate of the uncertainty in $dn/dz$, as it neglects the correlations 
in the errors of the Schechter parameters.

\begin{center}
{TABLE 1\\[4pt] \scshape 
Median redshift $z_m$ in the four magnitude bins computed using three different methods}
\\[3pt]
\nopagebreak
\begin{tabular}{llll}
\tskip\tableline\tableline\tskip  Magnitude Bin & CNOC2 &
 CFRS+SDSS &  SDSS Photo-$z$  \\
\tskip\tableline\tskip\tskip
$18-19$
	& $0.17 \pm 0.014$	
	& 0.17		
	& 0.20	 	\\
$19-20$
	& $0.24 \pm 0.025$	
	& $0.25		$
	& $0.26	 	$\\
$20-21$
	& $0.33	\pm 0.037$
	& $0.35$		
	& $0.33$	 	\\
$21-22$
	& $0.43 \pm 0.062	$
	& $0.46$		
	&  	 	\\
\tableline
\end{tabular}
\end{center}

Figure \rf{dndz_all} plots a set  of $dn/dz$ from \ec{dndzpar} 
in the magnitude bin $20<\rstar<21$. 
The apparent discrepancy between the fit and the direct CNOC2 redshift 
has a negligible effect on the derived power spectrum.
We will show this by using both $dn/dz$'s when computing
the power spectrum; we will see that the differences in the derived power spectra 
are much smaller than the error bars 
(see Figure \rf{cnoc2} in \secc{invert}). 
Apparently, noticeably different radial galaxy distributions project the 3D 
clustering the same way on the sky as long as their median redshifts 
are the same.

\begin{figure}[thbp]
\centerline{\epsfxsize=5truein\epsffile{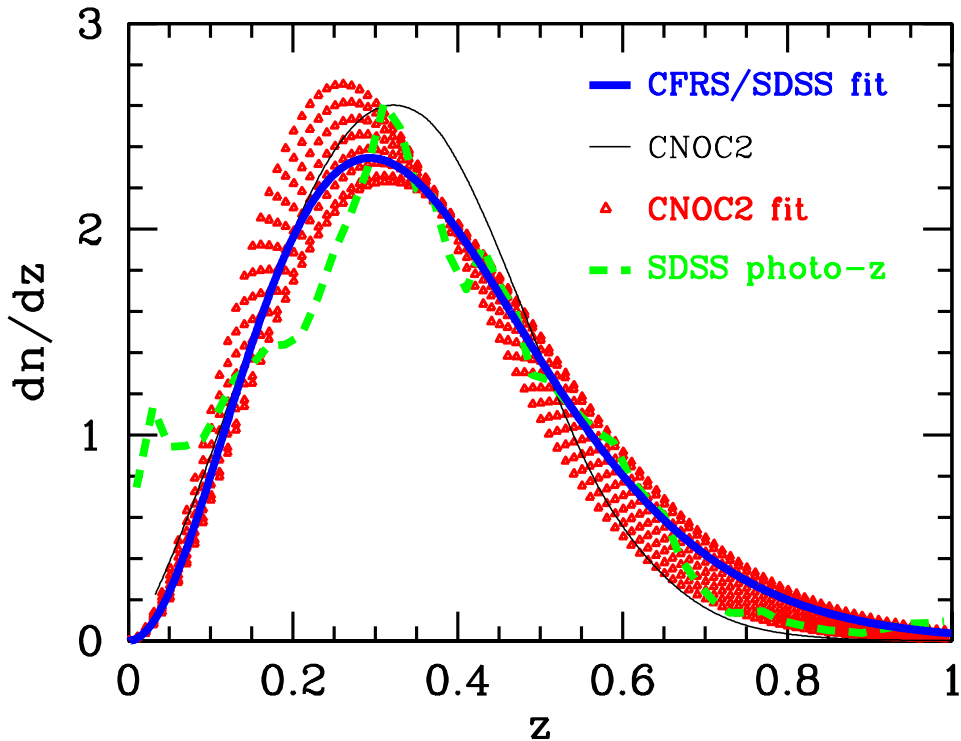}}
\caption{Redshift distributions in
the magnitude bin $20 < \rstar < 21$. Thin black curve is the direct CNOC2 result;
red triangles span the $1-\sigma$ range associated with the fit of 
\ec{dndzpar} with $z_m=0.33\pm 0.037$; thick blue curve
is best fit of \ec{dndzpar} to CFRS+SDSS data; dashed green curve uses the
photometric redshifts from SDSS.}
\label{fig:dndz_all}
\end{figure}

\subsection{CFRS+SDSS}

The Canada-France Redshift Survey (CFRS, Lilly et al. 1995)
is a deep redshift survey (with median redshift $z_m \approx 0.6$) 
of about $10^3$ faint galaxies, evenly subsampled 
in the magnitude range $17.5 < I(AB) < 22.5$.
These galaxies were selected from five survey fields, each  of area 
$10' \times 10'$, and also have deep $V$ and $I$ band isophotal photometry.
The galaxy photometry was carried out using a limiting isophote of 
$\mu_{AB}(I) = 28\ $mag$\ $arcsec$^{-2}$, and therefore would be expected
to contain almost all  of the light.
We used an empirical relation $r_{\rm SDSS} =  I_{\rm CFRS} + 0.5*(V-I)_{\rm CFRS}$ 
derived using 130 galaxies that were imaged by both SDSS and CFRS, 
to estimate the $\rstar$ band magnitude of all the CFRS galaxies that
also have redshifts.
We fit a third order polynomial for the median redshifts of the galaxy 
distribution as a function of $\rstar$ $[z_m(\rstar)]$ in the magnitude range
$17 < \rstar < 22$, using the SDSS redshifts in the magnitude range
$17 < \rstar < 17.6$ and the CFRS redshifts at $19 < \rstar < 22$.
Finally, we determined the median redshift of galaxies in each
unit magnitude bin in the magnitude range $18 < \rstar < 22$ by
multiplying this $z_m(\rstar)$ with a third order polynomial fit to
the galaxy number counts as a function of magnitude in the SDSS
(Yasuda et al. 2001) in the magnitude range $18 < \rstar < 22$.
The median redshifts in the four unit magnitude bins are given in Table 1;
they agree remarkably well with the CNOC2 estimates. The agreement is 
impressive especially since each data set has its own limitations. 
While CFRS is a deeper survey over a smaller area, CNOC2 has no redshifts 
greater than $0.55$ (more precisely, their Schechter fits are valid only
for $z<0.55$ since their efficiency falls steeply beyond this redshift).
The fit in the magnitude bin $20-21$, shown in Figure \rf{dndz_all},
is consistent with the CNOC2 results. 

\subsection{Photometric Redshifts}

Redshifts for the SDSS photometric data were estimated directly from
the $u^*, g^*, \rstar, i^*$ and $z^*$ galaxy photometry. The technique used in
this analysis is a hybrid between the empirical approach of fitting
a polynomial relation between the colors and redshifts of galaxies
(Connolly et al 1995) and techniques that model the colors of
galaxies as a function of redshift using {a priori} galaxy spectral energy
distributions (Fernandez-Soto et al 1999, Sawicki et al 1997). In our
approach, we compare the colors of the galaxies with those predicted
from galaxy spectral templates but we construct these templates from a
sample of galaxies with high signal-to-noise photometry and
spectroscopic redshifts (Connolly et al. in preparation).
In this way, we can construct templates that
are designed to match the observed colors of galaxies within a particular data
set. The training techniques employed in this analysis are described
by Csabai et al (2000) and Budavari et al (2000). The training of the
spectral templates was based on $5,000$ SDSS galaxies (with measured
redshifts) supplemented with $700$ galaxies with redshifts from the
CNOC2 survey, for which we have SDSS photometry. For
the trained spectral templates the dispersion within the photometric
redshift relation is $\sigma_z \approx 0.07$ for $\rstar<19$ and $\sigma_z
\approx 0.12$ for $\rstar<21$. All galaxies with photometric redshifts have
associated errors and spectral types.

The redshift distribution from photo-$z$ is shown in the magnitude bin $20-21$ 
in Figure \rf{dndz_all}. In the regime of interest it agrees well with other 
estimates; the small fluctuations at $z < 0.2$ probably arises from
large scale structure in the galaxy distribution.
Table 1 presents the median redshift in all magnitude bins
except the faintest bin, where we are limited by the uncertainties
in evolution of the galaxy population.

\section{Kernels}
\label{sec:kernels}

The angular correlation functions in real and Fourier
space are related to the 3D power spectrum (e.g. Peebles 1980; Eisenstein
\& Zaldarriaga 2000) by
\be
 \left( \matrix{w(\theta)\cr l^2 C_l}\right)
 = \int_0^\infty dk\,k\,P(k,z=z_m) \left( \matrix{
K_w(k\theta) \cr
K_{C_l}(k/l)}\right)
\eql{kernels}
\ee
in the small angle approximation. 
Here, $z_m$ is the median redshift of the galaxy distribution in 
a given magnitude bin in the angular photometric catalog. The kernel
for $w$ is
\be
 K_w(k\theta) = {1\over 2\pi} 
\int_0^\infty dz (d\chi/dz) J_0(k\theta \chi(z))
\left[(dn/dz) (dz/d\chi) \right]^2 E(z,z_m)	
.\ee
Here, $\chi$ is the comoving distance out to redshift
$z$; $J_0$ is the Bessel function of order zero; $dn/dz$ is the
redshift distribution of galaxies, normalized so that the integral
of $dn/dz$ over all $z$ is one; and $E(z,z_m)$ is a function
which describes the redshift evolution of density fluctuations between
the redshifts $z_m$ and $z$.
In principle $E$ is quite complicated, with a scale ($k$) dependence as well.
However, for all practical purposes the current data are not sensitive to 
this evolution (Scranton \& Dodelson 2000), so we set $E=1$. 
In other words, the angular clustering in a given magnitude bin is determined
by the 3D clustering at the median redshift associated with that bin.
Thus, when we invert the angular correlations to obtain the 3D power spectrum,
we should keep in mind that the different spectra obtained from the different
magnitude bins correspond to different redshifts.
We assume a flat, $\Lambda$ dominated Universe with $\Omega_\Lambda=0.7$.
The kernel $K_w(k\theta)$ is shown in Figure \rf{kernel} for the four 
magnitude bins. The largest contribution to $w(\theta)$ comes from scales 
on the order of the first zero, e.g. for $21 < \rstar < 22$, 
at $k\sim 0.002 {\rm h \ Mpc}^{-1}/\theta$, and a factor of three smaller
scale at $k\sim 0.006 {\rm h \ Mpc}^{-1}/\theta$ in the $18 < \rstar < 19$ bin. 
Thus, as anticipated, the angular correlations of galaxies at a given 
angle in the faint magnitude bins probe larger scales (smaller $k$) than 
do those in the bright magnitude bins.

\begin{figure}[thp]
\centerline{\epsfxsize=5truein\epsffile{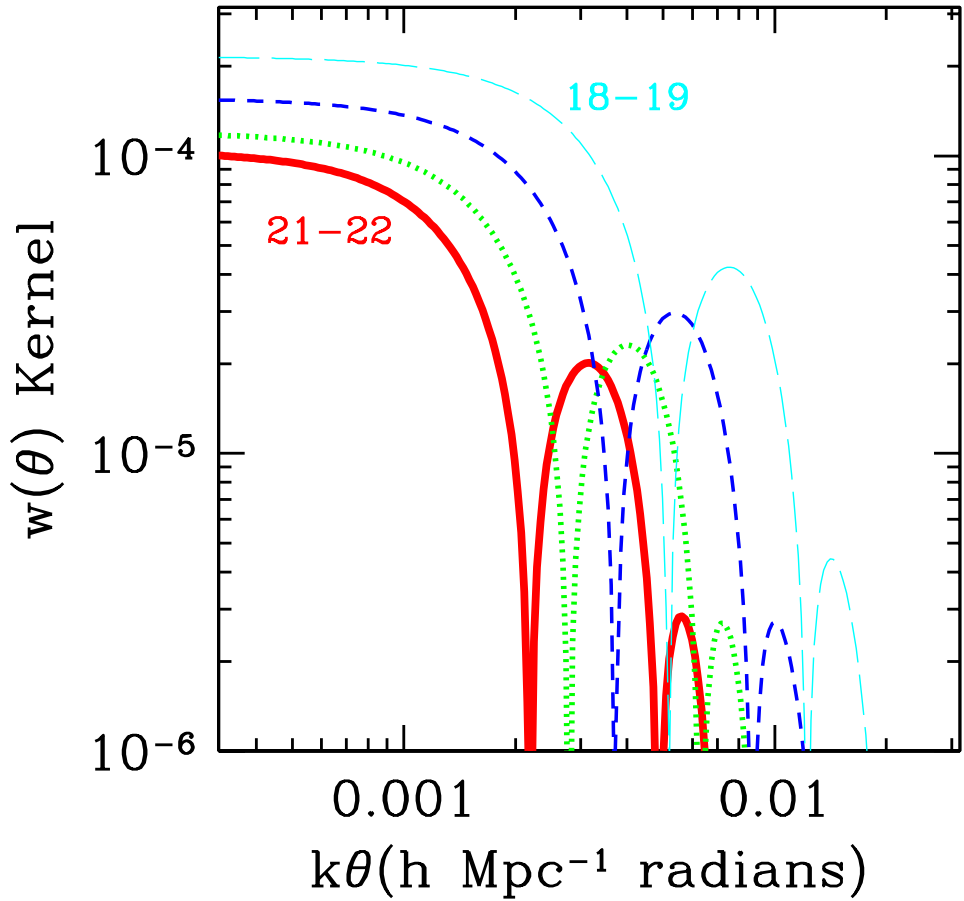}}
\caption{Absolute value of the kernel for the angular correlation 
function $w(\theta)$ in four different magnitude bins. At left, faintest bin
($21 < \rstar < 22$) is lowest, up to the brightest bin 
($18 < \rstar < 19$) on top. The sign of the kernel starts positive at small 
$k$ and changes with each oscillation.}
\label{fig:kernel}
\end{figure}

The kernel for $C_l$ is simpler (again, in the small-angle approximation)
\be
K_{C_l}(k/l) = \left[ \chi E(z,z_m) \Big( {dn/dz) (dz/d\chi)} 
\Big)^2
\right]_{\chi=l/k}
.\ee
Both of these kernels $K_w$ and $K_{C_l}$ should be generalized to large
angles, once the angular correlations are measured on large angular scales.
However, the current measurements of angular clustering in the galaxy
distribution are on small enough scales that these small angle 
expressions remain valid (see e.g. T01).
Figure \rf{clkernel} shows the kernels $K_{C_l}$ in the
four magnitude bins. Here, it is easier to see (as compared with
the $w$ kernels) that (a) the kernel $K_{C_l}$ has a single peak
showing that the $C_l$ on a given angular scale gets most of its 
contribution from the 3-dimensional correlations on a given physical 
scale, and (b) this physical scale is larger for fainter magnitude bins 
(since the peak moves to the left) than for brighter bins.  
Another feature which makes the $C_l$'s easier to interpret is that the 
kernel is always positive.

\begin{figure}[thp]
\centerline{\epsfxsize=5truein\epsffile{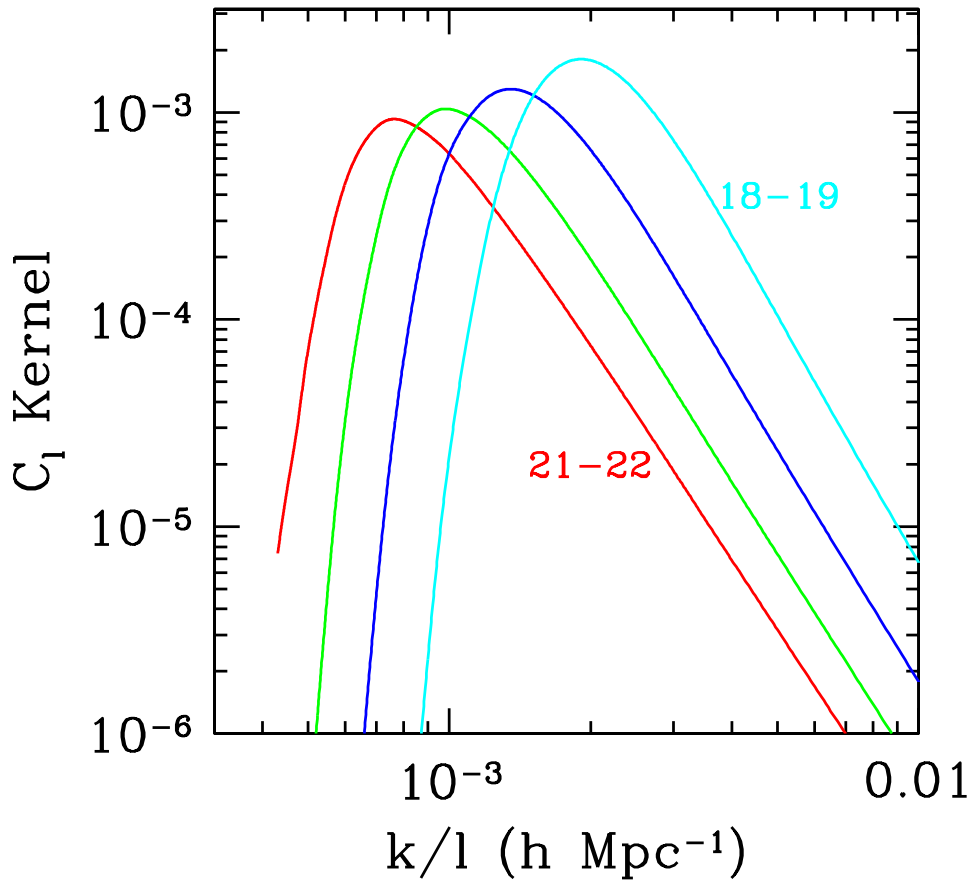}}
\caption{Kernel for the 2D power spectrum ($K_{C_l})$ in four magnitude
bins. The kernel for galaxies in the faintest magnitude bin is the one to
the farthest left, 
since faint galaxies probe larger physical scales at a fixed angular scale.}
\label{fig:clkernel}
\end{figure}

The uncertainty in $dn/dz$ leads to a corresponding uncertainty in the kernels.
It is most instructive to consider this resulting uncertainty in the $C_l$ 
kernel since it is a relatively simple function. 
Figure \rf{kernel_spread} shows the level of this uncertainty in $K_{C_l}$. 
Using the parameterization of \ec{dndzpar}, we can change the
distribution by varying the parameter $z_m$. When $z_m$ is small, a given 
angular scale subtends a small physical scale. Hence the kernel peaks at 
large $k/l$ (for fixed $l$, the wavenumber $k$ is large). As $z_m$ increases, 
a given angular scale subtends a larger physical scale. We therefore see the 
kernel shift to the left in Figure \rf{kernel_spread}. Besides this lateral 
shift, the amplitude decreases as $z_m$ increases. To get the same angular 
clustering strength, one needs a larger 3D power spectrum as the median 
redshift of the galaxies increases to avoid cancellations of over- and 
under-densities along the line of sight. Also shown in 
Figure \rf{kernel_spread} is the shift in $K_{C_l}$ arising from a different
background cosmology. The peak of the kernel in a flat, matter-dominated
Universe shifts towards smaller scales. This is to be expected, as 
a flat $\Lambda-$dominated  Universe has ``more volume'' at a given redshift.
Therefore, the physical distance separating galaxies at a given $z_m$ is 
larger in a $\Lambda-$Universe. Clustering on a given angular scale therefore
probes larger physical scales (smaller $k$) in a $\Lambda-$Universe.
The amplitude of the kernel decreases as $\Lambda$ is introduced
because there are more cancellations
in the density inhomogeneities along the line of sight.

\begin{figure}[thp]
\centerline{\epsfxsize=5truein\epsffile{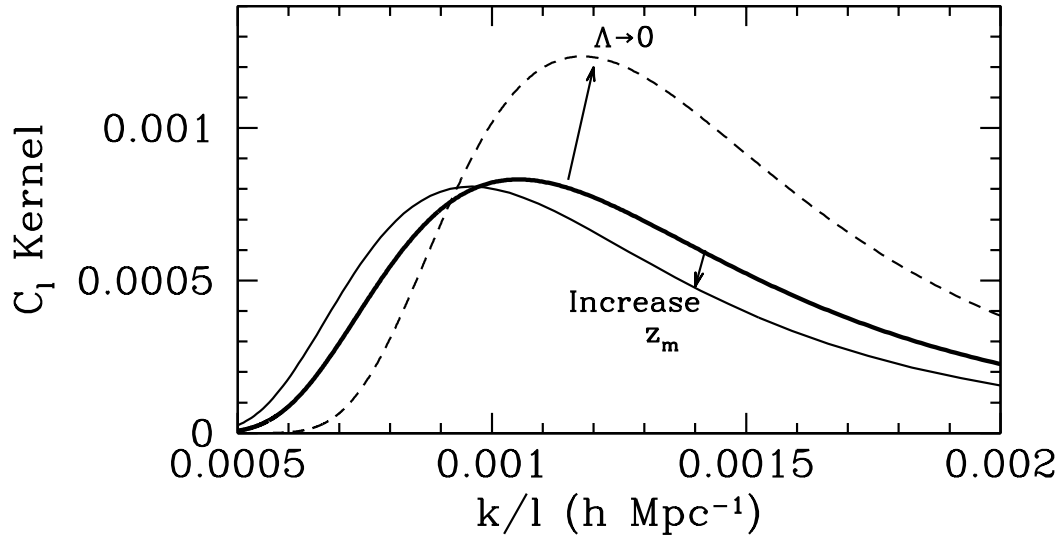}}
\caption{The sensitivity of the angular kernel 
(in the magnitude bin $20 < \rstar < 21$)
to uncertainties in the redshift distribution and cosmology. Heavy solid
(red) curve corresponds to the best fit of \ec{dndzpar}, with $z_m=0.33$. 
Thin (red) curve shows the effect of increasing $z_m$ by $1-\sigma$.
Dashed (blue) curve shows the effect of modifying the underlying cosmology,
and corresponds to the kernel in a flat, matter-dominated Universe.
It has the same $z_m$ as the heavy red curve.}
\label{fig:kernel_spread}
\end{figure}

\section{The 3D Power Spectrum}
\label{sec:invert}

The Automated Plate Machine (APM; Maddox et al. 1990) survey  a decade ago 
ushered in a renewed interest in techniques for inverting the angular 
correlation function to extract the 3D power spectrum. 
Efstathiou and Baugh (1993) used Lucy's Method to accomplish this inversion; 
Dodelson and Gaztanaga (2000) used a Bayesian prior constraining the
smoothness of the power spectrum to stabilize the inversion; 
most recently, Eisenstein and Zaldarriaga (2000, hereafter EZ) used a 
Singular Value Decomposition to eliminate modes that destabilize the 
inversion. 

The good news is that all these techniques ``work.''
They have been tested using numerical simulations
and against each other and generally they get similar results for the
power spectrum. The more subtle issue has been a correct estimate of the
errors on the derived power spectrum.
Perhaps most important, as mentioned by Dodelson and Gaztanaga (2000) and 
then illustrated persuasively by EZ, errors on $w(\theta)$ measurements
in different angular bins are highly correlated. 
EZ showed that accounting for these correlations is crucial if one wants 
to report reliable errors bars on the 3D power spectrum.
This flurry of activity, along with parallel activity in map-making
(another form of inversion in the field of Cosmic Microwave Background
anisotropies; see e.g. Bond et al. 1999 for an overview) has taken the 
inversion process to a higher level of sophistication and reliability. 

The starting point for inversion is the relation between the measured
angular clustering of galaxies and the 3D power spectrum, the main component 
of which is the kernel. The integral in \ec{kernels}
can be written in matrix form as
\be
d = \matk P + n
\eql{dsn}
\ee
where $d$ is an $N_d$ dimensional vector containing the data, 
either the angular correlation function 
$w(\theta)$ (in which case each component of the vector corresponds 
to $w$ in a given angular bin) or the 
2D power spectrum $C_l$ (with each component the power in an $l-$bin). The 
kernel $\matk$ is an $N_d\times N_P$ matrix, $N_P$ being the number of bins in 3D wave 
number $k$ in which we are attempting to estimate the power spectrum $P$.
Note that $\matk$ is related to, but not identical to the kernels of
\ec{kernels}, due to the extra factor of $k\,dk$.
In addition to the signal (the first term in \ec{dsn}), the data
points (measurements) also contain some noise, characterized by the
noise vector $n$. 
We take this noise to be drawn from a distribution with mean zero and 
variance given by the covariance matrix, $C_d$.
The covariance matrix for $w$ is discussed in detail in Sc01, that for
$C_l$ in T01.

Following EZ, we re-define the data and power spectrum vectors. 
The data is renormalized by the noise:
\be
d' \equiv C_d^{-1/2} d
\ee
where $C_d^{-1/2}$ is the matrix with the property 
$C_d^{-1/2} C_d^{-1/2} = C_d^{-1}$. 
Roughly, $d'$ is the signal to noise ratio of the actual data points 
(this is easiest to see in the unrealistic case that the correlation
matrix of the noise is diagonal).
The power spectrum is also renormalized, weighted by some fiducial spectrum 
$P_{\rm norm}$:
\be
P' \equiv {P \over P_{\rm norm} }.
\ee
With this renormalization, all the elements of the vector $P'$ will
be of order unity; large deviations can be flagged.
We set the fiducial spectrum to be that of a $\Lambda CDM$ model,
with parameters identical to those used in the simulations
described in Sc01 (e.g. $\Omega_\Lambda=0.7$).
although our final results are insensitive to this choice. 
In terms of these renormalized variables, the relation
between data and signal (\ec{dsn}) becomes
\be
d' = \matk' P' + n'
\ee
with the renormalized kernel equal to
\be
\matk' = C_{d}^{-1/2} \matk P_{\rm norm} 
.\ee
The renormalized noise $n' = C_{d}^{-1/2}n$ has covariance
matrix equal to the identity matrix 
(since $<nn>=C_d$).
The new kernel $\matk'$ is roughly  the expected signal ($\matk P_{\rm norm}$) 
divided by the noise. Therefore, modes with eigenvalues of $\matk'$ greater 
than one have large signal, while modes with very low weight are not 
expected to yield much information on the power spectrum.

With these redefinitions, the minimum variance estimator of the power spectrum
is
\be
\hat{P'} = C_{P'} \matk^{'t} d'
\eql{pest}\ee
with the associated error matrix
\be
C_{P'} = \left( \matk^{'t} \matk'\right)^{-1}
,\eql{noise}\ee
where $^t$ means transpose. To get back to $P$ and its covariance matrix, 
one simply multiplies each element of $\hat{P'}$ by the corresponding element 
of $P_{\rm norm}$ and each element of the matrix $C_{P',ij}$ by 
$P_{{\rm norm},i}P_{{\rm norm},j}$. If we include all the modes associated 
with the rows of $\matk'$, then $P_{\rm norm}$ has absolutely no effect on the 
inversion. Only when we begin eliminating modes with low signal-to-noise
ratio (as described below) does the choice of $P_{\rm norm}$ have any impact.

In practice, the estimator of \ec{pest} is not always stable: very small
changes in the input data lead to huge variations some elements of $P'$.
EZ (and other groups making maps from Cosmic Microwave Background timestreams)
showed that an easy way to handle this instability is to use Singular Value 
Decomposition, wherein the matrix $\matk'$ is decomposed as
\be
\matk' = U W V^t
\ee
where $U$ is a column orthonormal matrix, $W$ is a diagonal matrix,
and $V$ is orthogonal. The weights, the diagonal elements of $W$, then
are measures of signal-to-noise ratio. Very small
weights are responsible for the instability of the inversion. 

One form of instability is particularly easy to treat. If the eigenmode
with a very low weight can be identified with a single $k$ bin, then
we eliminate the low weight by simply removing the $k$ bin. We then redo
the inversion without the offending bin.
Especially in the case of $w(\theta)$ inversion we will see that this 
way of tweaking the choice 
of $k$ bins eliminates much of the instability.
An example of this process is illustrated in \secc{invertw}. 

As we will see below, this improved choice of $k$ bins
does not work as well on the $C_l$ data of T01. In that case, the
low weight modes are typically linear combinations of many different modes, 
so it is impossible to identify one offending $k$ bin. 
EZ give a recipe for eliminating the 
deleterious effects of these low weights, which involves rescaling them 
higher when computing the covariance matrix $C_{P'}$. We follow this recipe
here, eliminating modes with weight less than unity (note that this is where
our choice of fiducial power spectrum enters into the analysis). Since 
the weight is a measure of signal to noise, we are effectively eliminating from
the analysis the noisiest modes. This should not have any impact on the
final parameter determination.

\subsection{Inversion of $w(\theta)$}
\secl{invertw}

We use the measurements of $w(\theta)$ and its covariance matrix as 
presented in Sc01 and C01. We use the measurements on 
angles larger than $0.01$ degrees, leaving us with $19$ angular bins.

\begin{figure}[thp]
\centerline{\epsfxsize=4.5truein\epsffile{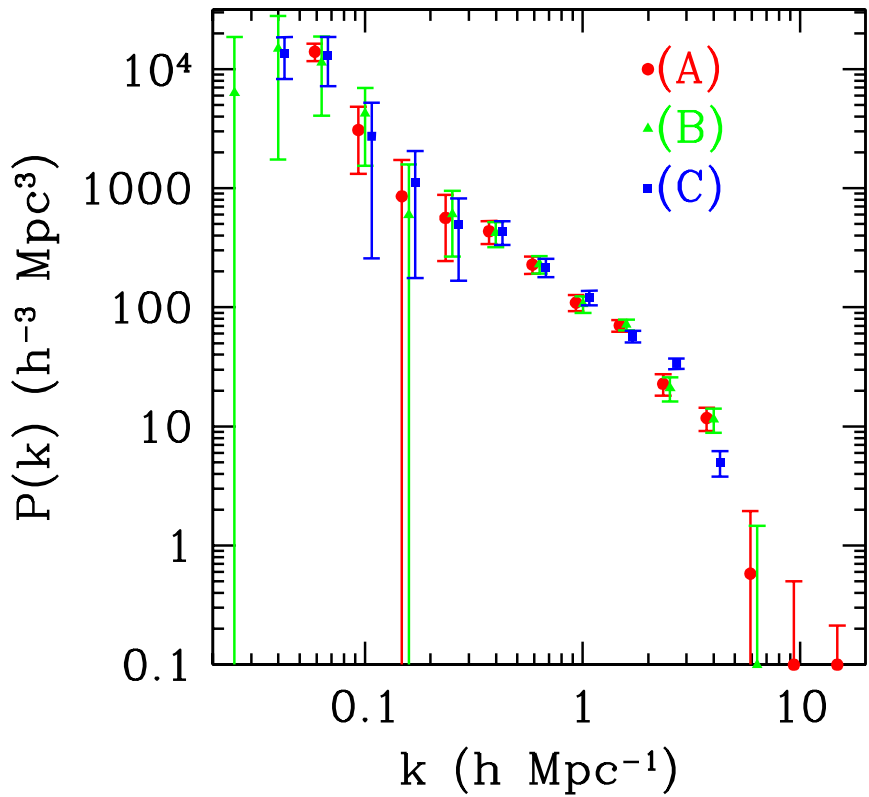}}
\caption{The power spectrum inferred from the bin $21 < \rstar < 22$.
The three different results correspond to three different choices of 
$k$ bins. 
First attempt (A) is the set depicted with red circles; second (B),
adding two low $k$ modes and dropping two high $k$ modes, with green 
triangles; final (C), dropping a $k$ mode on each end, with blue squares.
}
\label{fig:binning}
\end{figure}

Let us illustrate the Singular Value Decomposition with the $21 < \rstar < 22$
magnitude bin. For this bin, the kernel in Figure \rf{kernel} starts to 
fall off when $k\theta \sim 2\times 10^{-3}$h Mpc$^{-1}$ radians.
For a fixed angle $\theta$, the integral over $k$ in Eq.~[6] is dominated
by the value of the kernel in this region. Therefore,
the smallest angular bin considered here, $\theta = 0.01^\circ$,
is sensitive to the power spectrum around the 
wavenumber $k\sim 2\times 10^{-3}$h Mpc$^{-1}/1.7\times 10^{-4}
\sim 10$h Mpc$^{-1}$. 
Similarly, the largest angular scale with a significant measurement
of $w(\theta)$, around one degree, probes the power spectrum at 
wavenumbers of order $k \sim 0.1$h Mpc$^{-1}$.
Our first choice of $k$ bins with which to fit the power spectrum, 
therefore, spans the range\footnote{The bin at $k_i$ includes the range
$(k_{i-1}+k_i)/2 < k < (k_{i+1} + k_i)/2$, although, as we discuss
later in this section, the true {\it window function} is not a tophat.}
$k = 0.1-10$h Mpc$^{-1}$. 
As shown in Figure \rf{binning}, we actually try bins with a slightly 
broader range. The initial choice, labeled (A) in  Figure \rf{binning}, 
goes as low as $0.06 $h Mpc$^{-1}$ and as high as $15$h Mpc$^{-1}$. 
The results shown with this set of bins are after the rescaling of the
two low weights. We wish to avoid such rescaling though, so let us 
try to understand the origin of the low weights.

Figure \rf{w2122} shows the eigenvectors corresponding to thirteen
different weights of binning scheme (A), 
and their sensitivity to wavenumber. 
The two modes with largest weights (at the top) measure linear combinations 
of the power in the range $k \sim 0.1 \rightarrow 1\ $h Mpc$^{-1}$. 
The next six modes also probe this region, but these modes also
extend to smaller and larger scales. Note, though, that the ten modes
with the highest weights have little
sensitivity to the power spectrum in the last two bins,
at $k=8{\rm\ and}\ 15h$ Mpc$^{-1}$. They are however sensitive to the largest
scales in this binning scheme. 
We can also see this from the small error bars on the left-most red point 
in Figure \rf{binning}.
It is therefore natural to add a couple of bins at smaller $k$ and see 
whether any information can be obtained on very large scales. 

\begin{figure}[thp]
\centerline{\epsfxsize=6.5truein\epsffile{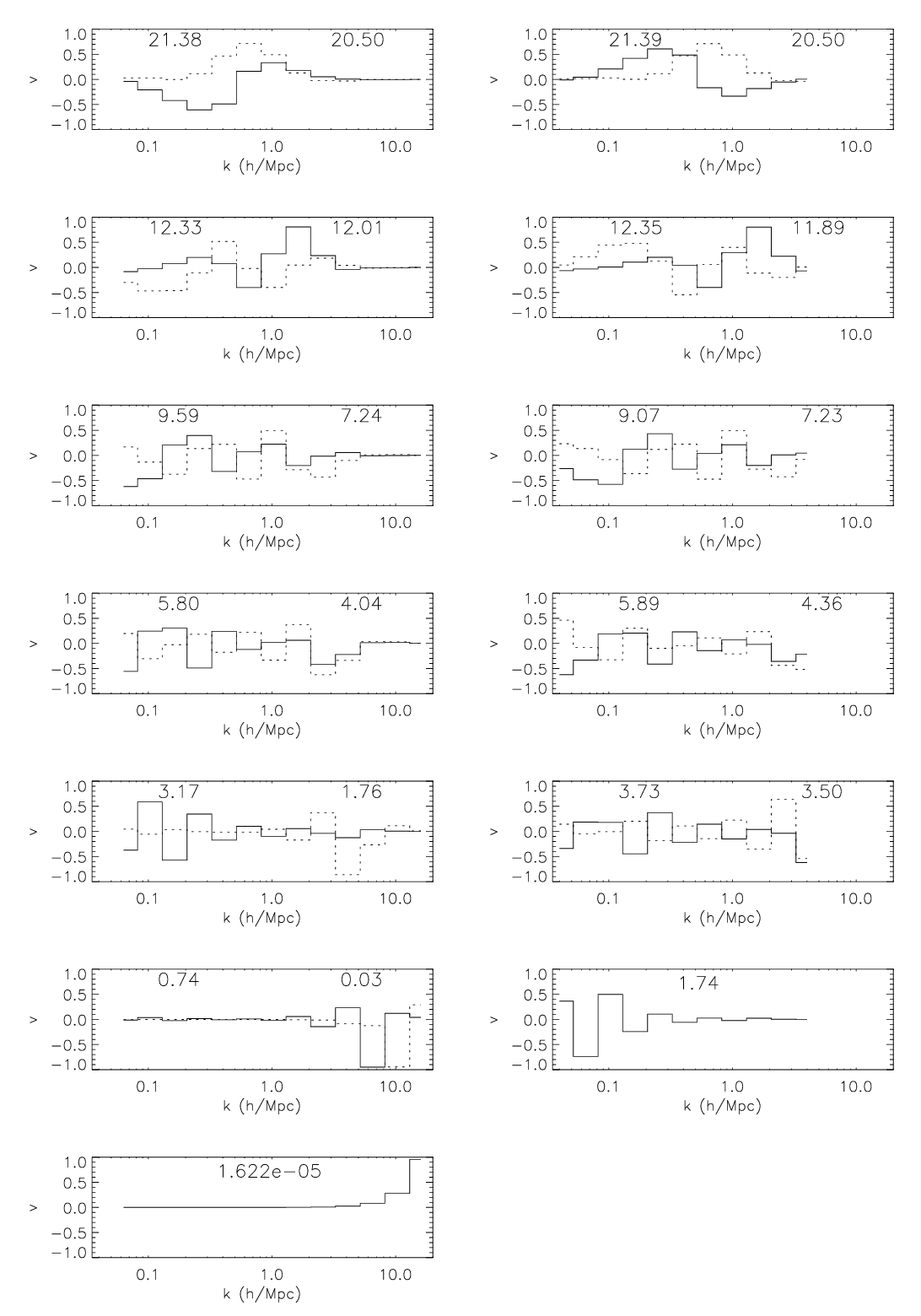}}
\caption{Eigenmodes of the renormalized kernel in the 
magnitude bin $21 < \rstar < 22$. Left column
shows all thirteen modes corresponding to $k$ binning scheme (A) in 
Figure \rf{binning}; right column shows all eleven modes corresponding to 
scheme (C). Modes are plotted two at a time for clarity, with the
dashed line corresponding to the mode with lower weight. The numbers
inside the panels are the weights of the two modes.}
\label{fig:w2122}
\end{figure}

The second scheme (B) in Figure \rf{binning} therefore adds two bins 
at $k=0.03,0.05$ h Mpc$^{-1}$ and drops the two highest $k$ bins, 
which we have identified with low weight modes. 
The measurement is indeed sensitive to the power in these two
large scale bins. It is also encouraging that the additional bins do not affect
the determination of the power anywhere else by more than about half 
a standard deviation.
It is clear then that our initial rough guess of the range to which 
$w(\theta)$ is sensitive was slightly incorrect: there is little sensitivity 
at the highest $k$ we chose, but we underestimated the large scale 
information. We still want to avoid rescaling weights, so we drop two more 
bins for our third and final binning scheme.
The right panels in Figure \rf{w2122} show the eigenvectors corresponding
to these eleven modes and their $k$ dependence. The highest weight modes are 
identical to those on the left (signs are irrelevant here).
All  modes have weights greater than unity; therefore the inversion is stable. 

It is important to make explicit the most important feature of Figure 
\rf{binning}, namely, the estimate of the power spectrum is robust to
reasonable changes in our choice of the $k$ bins.

\begin{figure}[htbp]
\centerline{\epsfxsize=4truein\epsffile{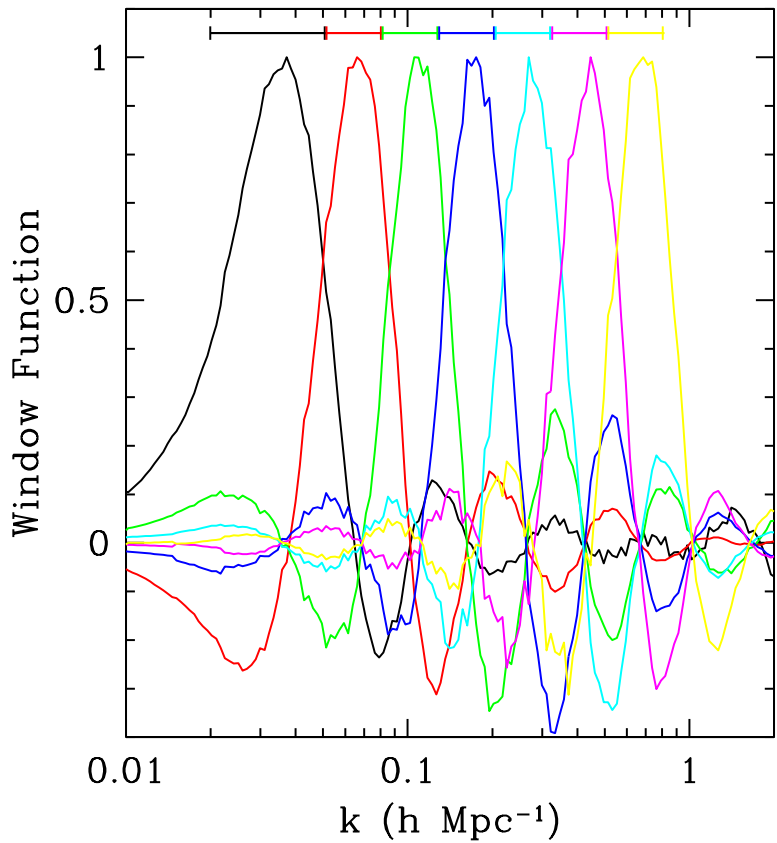}}
\caption{Window functions for the first seven $k$ modes in 
$21 < \rstar < 22$. The flat bands on top of each window function correspond to the
range of $k$ used in estimating the power in each mode. All the window
functions have been normalized to unity.}
\label{fig:winwth}
\end{figure}

Until now, we have been thinking of the power spectrum measured using \ec{pest}
as the best fit estimate of the power in a given band, assuming that the
true power spectrum was flat in that band. 
While this is correct, it is also not very useful, since realistic power
spectra are not flat within the finite width of the $k-$ bands of interest. 
Another way of understanding these spectra therefore is to note that the 
estimator in \ec{pest} is yet another in a series of quadratic estimators
(quadratic in the galaxy overdensities; see e.g.
Tegmark et al. 1998 and the discussion 
in Appendix A of T01). 
Any quadratic estimator has associated with it a window function
that relates it to the true power spectrum, and the one 
in \ec{pest} is no exception. In this case, the window function is
\be
\wind = C_{P'} K^{'t} K'_{\infty}
\eql{window}\ee
where $K'_{\infty}$ is $K'$ extended
to infinitely many  $k$ bins that are infinitesimally narrow.
In the limit that $K'$ is equal to $K'_\infty$, the window function
is simply equal to the identity matrix. The finite binning
changes this.
Some window functions for the faintest 
magnitude bin are shown in Figure \rf{winwth}. Note that they quite 
accurately probe the window for which they were constructed. The oscillating 
lobes of each window function remove power outside the band of interest.
These window functions must be used when comparing model power spectra
to the band power estimates presented here\footnote{The window functions along 
with all other information needed to compare models with the data 
(binned $P(k)$ and the associated covariance matrices) are available at  {\tt
http://www-astro-theory.fnal.gov/Personal/dodelson/sdss\_invert.html}}.

Let us now return to the question of ambiguity in the kernel arising
from the uncertainties in the galaxy redshift distributions. The fit of
\ec{dndzpar} to the direct CNOC2 $dn/dz$ appears only marginal in Figure 
\rf{dndz_all}. Figure \rf{cnoc2} compares the power spectrum derived
using the direct CNOC2 $dn/dz$ and the fit in equation \ec{dndzpar},
in the faintest magnitude bin $21 < \rstar < 22$.
The power spectra using these two $dn/dz$'s agree very well over the
entire range of wavenumbers, showing that our estimates of the power
spectrum are not very sensitive to the exact details of $dn/dz$.

\begin{figure}[thp]
\centerline{\epsfxsize=6.truein\epsffile{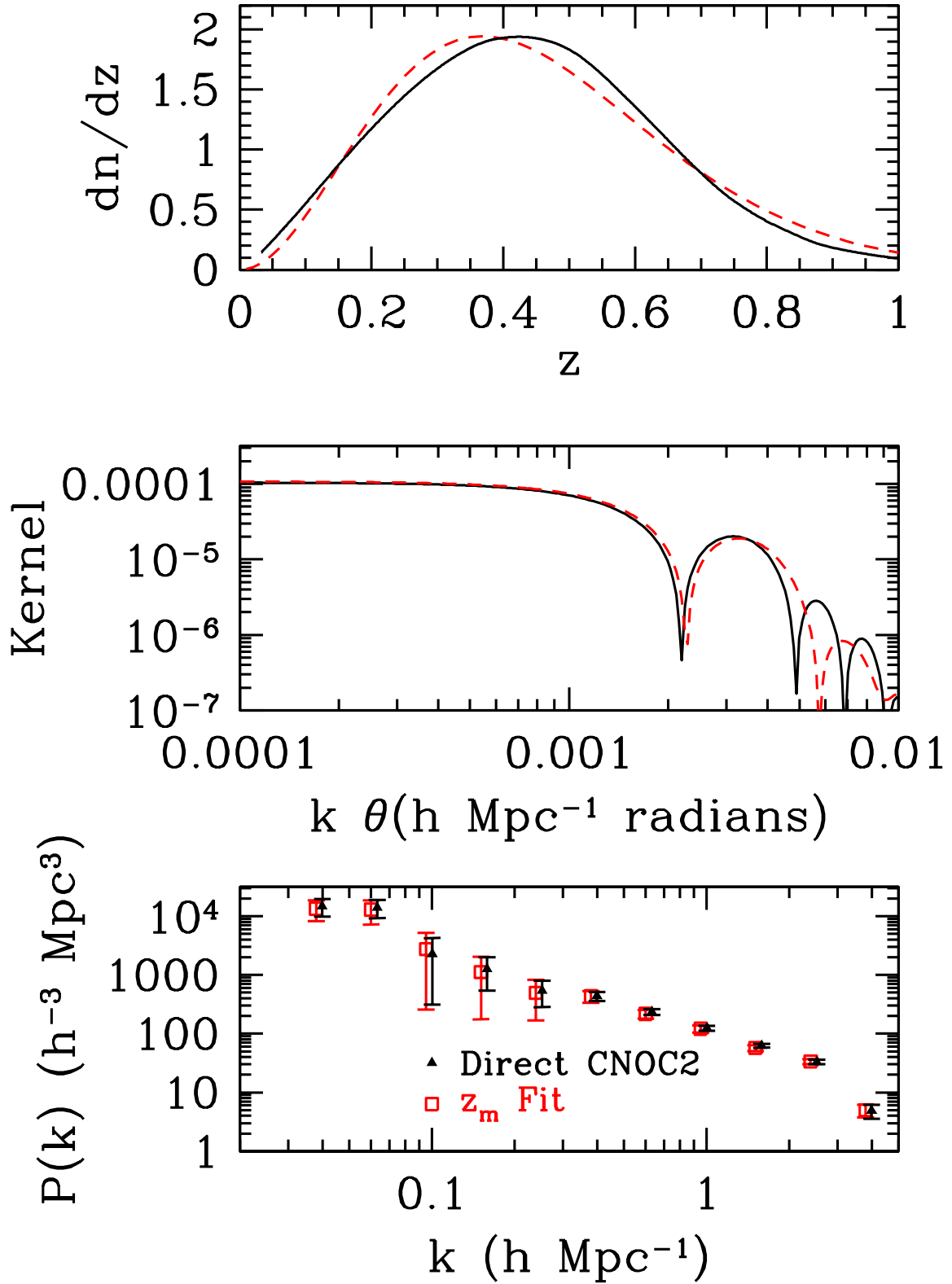}}
\caption{The result of fitting $dn/dz$ to the simple form of \ec{dndzpar}.
Black solid curves are the direct CNOC2 result, red dashed from \ec{dndzpar} with $z_m=0.43$
(all panels show $21 < \rstar < 22$). Top panel: Noticeable difference
in $dn/dz$. Middle panel: How this small difference affects the kernel. Bottom
panel: The final impact on the power spectrum is negligible.
}
\label{fig:cnoc2}
\end{figure}

The fit to \ec{dndzpar} therefore is adequate for our purposes. We can
use it to quantify the uncertainty in the galaxy redshift distribution.
In each magnitude bin, we account for the errors in $z_m$ and propagate 
these forward to estimate the errors on the power spectrum. To do this, 
we generate one hundred different $z_m$'s from a Gaussian distribution with
mean and standard deviation given in Table 1 (e.g. in the magnitude bin $21-22$, 
the mean is $0.43$ and standard deviation $0.062$). For each of these redshift 
distributions, we generate a kernel and compute the power spectrum from 
the $w(\theta)$ data. The spread in these hundred estimates of the
power spectra define a covariance matrix. This new covariance matrix is 
then added to the one defined in \ec{noise}.

\begin{figure}[thbp]
\centerline{\epsfxsize=5truein\epsffile{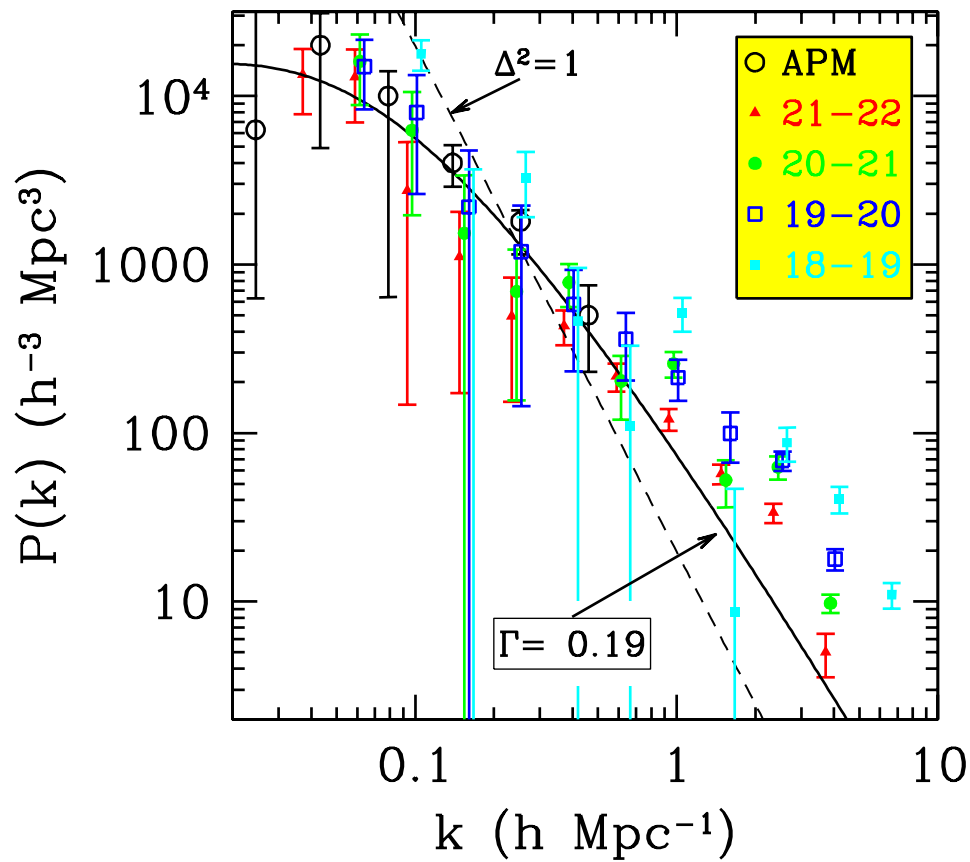}}
\caption{The 3D power spectrum from angular data in 
four magnitude bins. Also shown is the maximum likelihood power spectrum
from the APM survey (Efstathiou \& Moody 2000). The dashed line, defined
by $\Delta^2 \equiv 4\pi k^3 P/(2\pi)^3 =1$, delineates the
linear from the non-linear regime, while the solid line is the best fit 
linear spectrum from Sz01. The parameter $\Gamma$, introduced by
Bardeen et al. (1986) determines the shape
of the power spectrum.}
\label{fig:power_wtheta}
\end{figure}

Figure \rf{power_wtheta} shows the power spectrum from each of the four 
magnitude bins. Also shown in Figure \rf{power_wtheta} are the results
from the inversion of the APM angular data by Efstathiou and Moody (2000).
The two estimates of the power spectrum agree very well in the
overlapping wavenumbers $0.03 < k/{\rm h\ Mpc}^{-1} < 0.3$.
Figure \rf{power_wtheta} suggests that the power spectrum in the fainter
bins is smaller than in the brighter bins.
This is especially apparent for large $k$, in the non-linear regime. 
This realization is consistent with the fits of T01, which require larger
bias factors in the brighter bins.

There are several possible causes of this difference in power.
The first relies on the observation that the estimate from 
each magnitude bin samples the power 
at a different redshift. The median redshift for galaxies in the faintest 
bin is of order $z\simeq 0.43$, while the galaxies in the brightest magnitude
bin reside at redshift $z < 0.2$. Power in hierarchical models cascades
down from large to small scales as non-linearities become prominent. Therefore,
we expect less power at earlier times especially on non-linear scales\footnote{There is also the linear growth in power, but this effect is small for the 
redshifts here, especially in a $\Lambda-$model.}. Thus, consider the power 
spectrum in the $k=.252$h Mpc$^{-1}$ bin. The linear regime is usually 
associated with scales larger than those on which 
$\Delta^2 \equiv 4\pi k^3 P/(2\pi)^3 =1$. 
The two faintest magnitude points at $k=.252$h Mpc$^{-1}$
both have $\Delta^2$ significantly smaller than unity, while the APM point 
and that from our brightest magnitude are non-linear. Indeed, the usual 
cut-off between linear and non-linear scales lies in  the wavenumber
range $0.1 < k/{\rm h\ Mpc}^{-1} < 0.2$, so the linearity observed in our 
faintest magnitudes suggests we are indeed observing the Universe at 
earlier times when galaxies were less clustered.

Another possible cause of the difference in the power spectrum 
in different magnitude bins is the bias between
galaxies and mass. More luminous galaxies tend to be in clusters and 
therefore have a stronger power spectrum. This may be part of the explanation
for the larger power in the brighter magnitude bins.

Table 2 gives numerical values of the power spectrum in each of the four
magnitude bins, with two sets of errors. The first is the total error 
including the uncertainty in $dn/dz$, the second (in parentheses)
neglects this uncertainty.

\begin{center}
{TABLE 2\nopagebreak\\[4pt] \nopagebreak
\scshape 
3D Power spectrum in 4 magnitude bins.}
\nopagebreak
\\[3pt]
\nopagebreak
\begin{tabular}{lllll}
\tskip\tableline\tableline\tskip  k (h Mpc$^{-1}$)
& \multicolumn{4}{c}{P(k) (h$^{-3}$ Mpc$^{3}$) $\pm dP$ ($dP$ no $dn/dz$ uncertainty)}\\
\tskip\tableline 
&
 18-19 & 19-20
& 20-21 & 21-22 \\
\tskip\tableline\tskip\tskip
$ 0.040$ &             & &                   & $13409 \pm 5614 (5130) $\\
$  0.063$ &             & $14908\pm 6604 (4887)$   & $16016\pm 7168 (3507)$                  &$ 12916 \pm 5962 (5711) $\\
 $ 0.100$ & $17746\pm 3609 (3138)$          &  $7956\pm 5344 (5201)$   & $6246\pm 4292 (3142)$      &$ 2728\pm 2591 (2480)$\\
 $ 0.159$ & $-3692\pm 3664 (2929)$          &  $2191\pm 2569 (2483)$ & $1531\pm 1847 (1511)$        & $1112\pm 939 (936)$\\
 $ 0.252$ & $3274\pm  1371 (1187)$         &  $1191\pm 1047 (967)$  & $691\pm 535 (519)$            & $493\pm 340 (326)$\\
 $ 0.40$  &  $461\pm  492 (464)$        &   $580\pm 348 (338)$  & $782\pm 222 (191)$               & $432 \pm 101 (98)$\\
 $ 0.63$  &  $110\pm  218 (182)$          &   $360\pm 155 (127)$  & $203\pm 83 (69)$              & $217\pm 41 (38)$\\
 $ 1.00$  & $517\pm 117 (75)$           &            $213\pm 58 (56)$     & $258\pm 45 (30)$    & $121\pm 17.8 (16.7)$\\
 $ 1.59$  &$8.7\pm 38 (29)$             &         $99\pm 32.8 (20.9)$& $ 52.5 \pm 16.3 (15.5)$     & $57\pm 7.6 (6.4)$\\
 $ 2.52$  & $87.7\pm 19.9 (10.6)$            &          $68.6\pm 9.0 (6.9)$ & $62.9\pm 10.0 (5.8)$  & $33.7\pm 4.4 (3.4)$\\
 $ 4.0 $  & $40.6\pm 7.3 (4.1)$              &   $17.9\pm 2.6 (1.6)$ & $9.8\pm 1.2 (1.1)$                      & $5.0\pm 1.4 (1.2)$\\
$6.3$ & $11.0\pm     1.9  (1.2)$ & &                                                       &\\
\tableline
\end{tabular}
\end{center}

The errors on the 3D power spectrum estimates in different $k$ bins
are correlated. The error bars shown in  Figure \rf{power_wtheta} are the 
{\it marginalized} errors, allowing the power in all other $k$ bins to 
vary freely. These are the square roots of the diagonal elements of $C_{P}$. 
The {\it unmarginalized} errors (inverse square root of the diagonal 
elements of $C_{P}^{-1}$) are often much smaller due to the correlations.
Figure \rf{covar} illustrates the correlation matrix:
\be
r_{ij} \equiv  {  C_{P,ij} \over \sqrt{ C_{P,ii} C_{P,jj} }  }
\ee
for the power spectrum extracted from galaxies in the four bins. Typically,
adjacent $k$ bins are significantly anti-correlated. The correlation matrices
in Figure \rf{covar} do not include the contribution from the
uncertainty in $dn/dz$. Figure \rf{total_covar} shows the correlation 
matrices including this uncertainty. 
Note that, especially in the brighter bins, our estimate for the power on 
small scales is now coupled to the large scale estimate  because of 
the uncertainty in $dn/dz$.

\begin{figure}[htp]
\centerline{\epsfxsize=7truein\epsffile{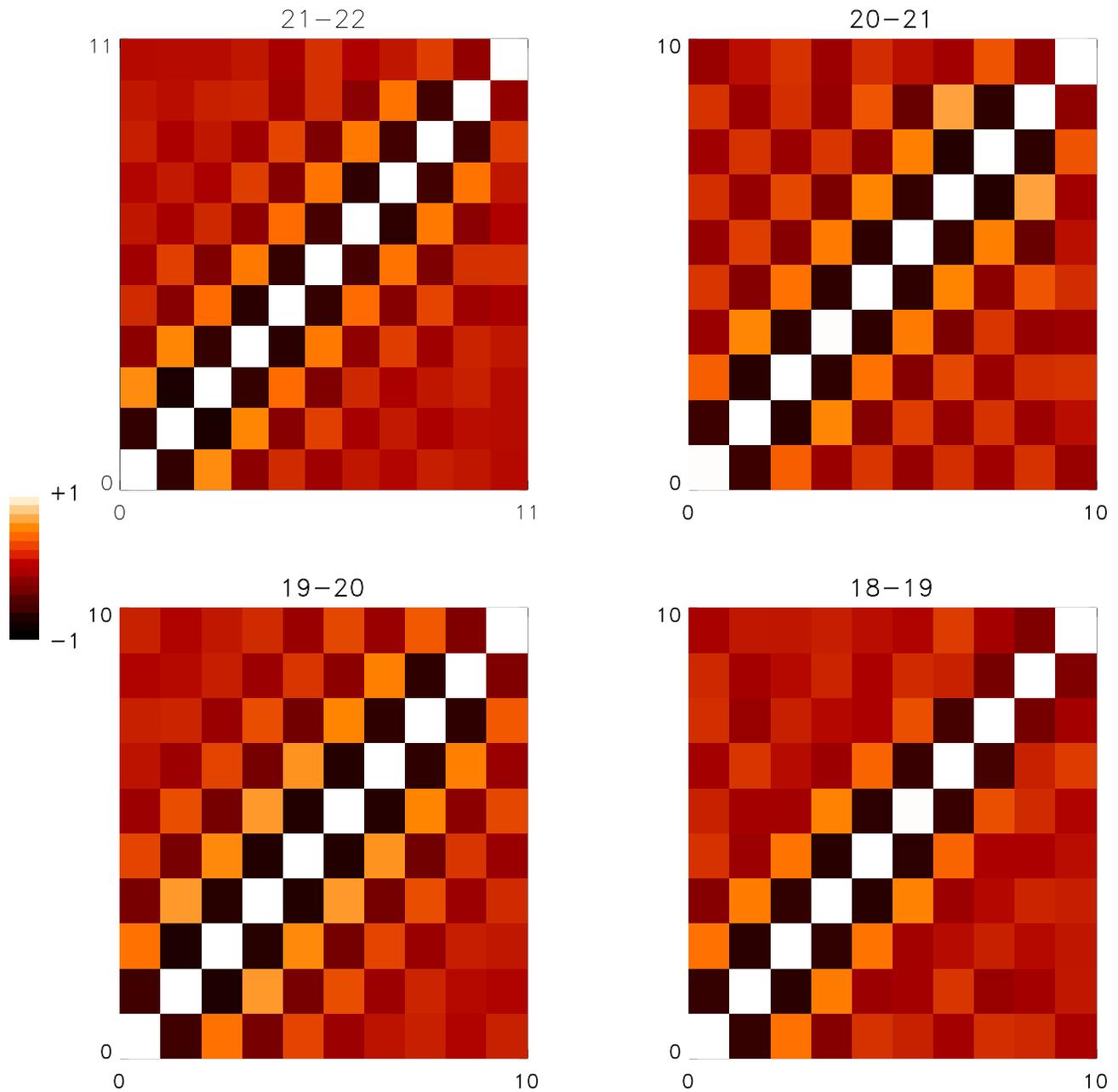}}
\caption{Correlation matrix for the 3D power spectrum inferred
from $w(\theta)$ in the four magnitude bins, neglecting the
contribution from the uncertainty in $dn/dz$.
Diagonal elements (all equal to one) run from bottom left to upper right. 
Elements in lower left are small $k$ (large scale) modes; e.g. bin labelled $0$
in upper left corresponds to $k=0.04$h Mpc$^{-1}$ as can be seen from Table 2. Adjacent $P(k)$ 
bins are anti-correlated at levels approaching $-1$.}
\label{fig:covar}
\end{figure}

\begin{figure}[htp]
\centerline{\epsfxsize=7truein\epsffile{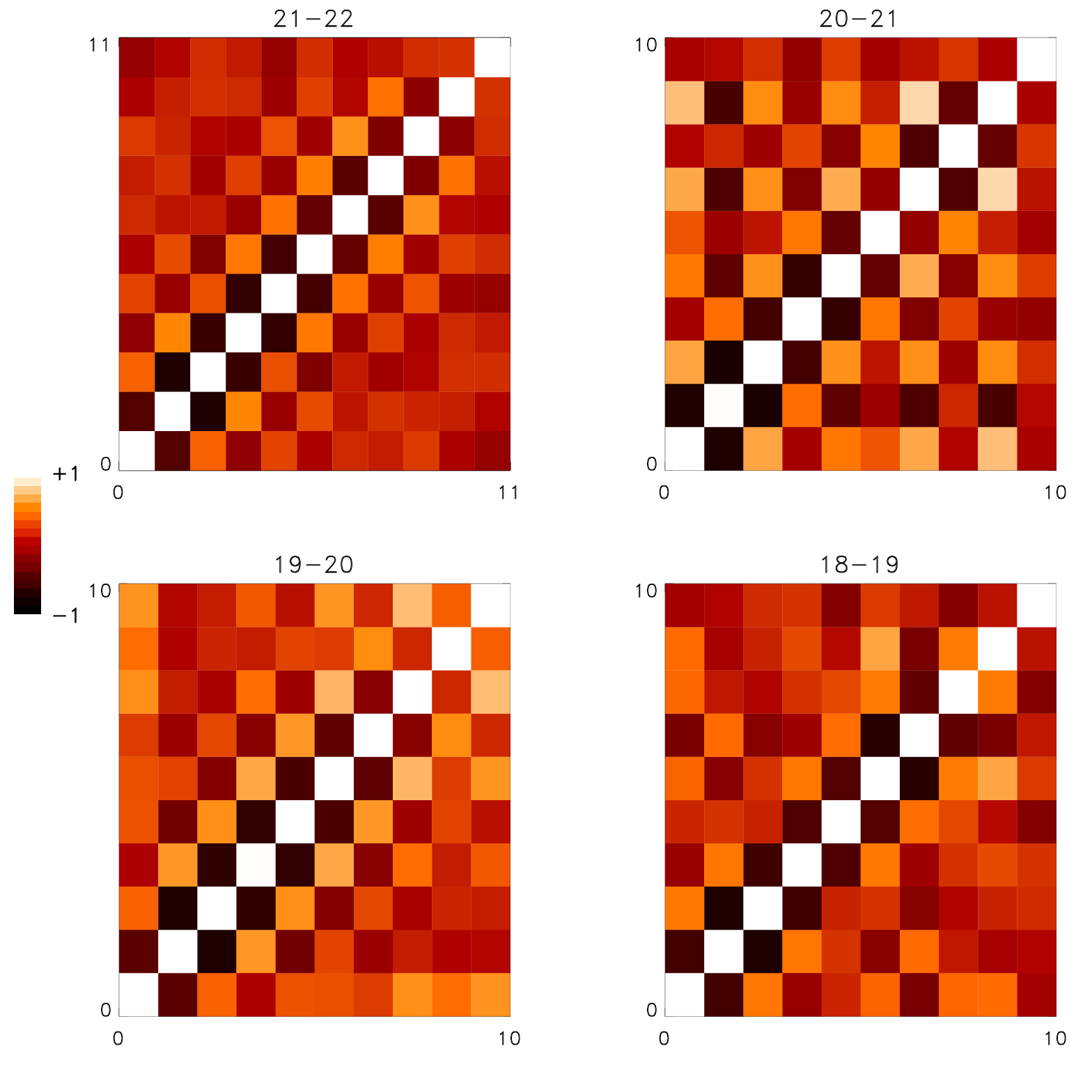}}
\caption{Same as Figure \rf{covar} except that the correlation matrix
includes the contribution from the uncertainty in $dn/dz$.
This leads to enhanced correlations between small and large scales.}
\label{fig:total_covar}
\end{figure}

\subsection{Inversion from $C_l$'s}

We now estimate the 3D power spectrum from the angular power spectrum,
the $C_l$'s, as presented in T01. The advantage of the inversion process
in this case is more subtle than in the case of $w(\theta)$. 
To understand why, recall that inverting is equivalent 
to introducing the quadratic estimator of \ec{pest}. As we have seen, this
estimator (\ec{pest}) has a window function associated with it. 
Looking back to \ec{dsn}, though,
we see that $C_l$ (or $w(\theta)$) itself is also a quadratic estimator 
of the power spectrum, with the window function equal to the kernel, $K'$. Let's
call this window function the {\it observational} window function to distinguish
it from that of \ec{window}.
The observational window function for $w(\theta)$ is not positive definite 
and is not sharply peaked around
a particular wavenumber (recall Figure \rf{kernel}). The observational window function
 for $C_l$ on the other 
hand does suggest that each $l-$ band can be associated with a given range in $k$.
So $C_l$ itself is a useful estimator of the power spectrum. Why invert then?
Inversion is still important, even for the $C_l$'s, because the window functions
associated with the estimator of \ec{pest} are more sharply peaked than the
observational window functions.

Figure \rf{winmax} illustrates this point. For three different $k$ modes, 
the window functions associated with \ec{pest} are plotted
along with the observational window functions. The estimator of \ec{pest}
tries to remove all power outside the bin of interest by introducing 
modulations of the power. The power estimated through the inversion process 
therefore gets very little leakage from higher or lower wavelength modes. 
This contrasts with the observational window functions, which are relatively broad and do 
pick up power from a wide range of scales. This feature is particularly 
important if one is interested in using the power spectrum estimates for 
parameter determination. In that case, one typically wants to restrict the 
analysis to the largest scales, which are least affected by non-linearities
and scale dependent galaxy bias. The narrow window functions arising from 
inversion facilitate  this restriction. 

We fit the fourteen measurements of $C_l$ with fourteen $k$ bands.
This binning allows for easy comparison with the observational
window functions, but it leads to more instability in the inversion
(many modes are undetermined).
This shows up in Figure \rf{winmax},  
where the lowest $k$ window has been shifted over significantly to 
smaller scales 
(larger $k$) as compared with the corresponding observational window. 
This shift is due to the elimination of low weight modes in the SVD. 
Since the largest scale modes have very little weight, the effective 
window moves to smaller scales that are better probed by the angular data.

\begin{figure}[htp]
\centerline{\epsfxsize=5truein\epsffile{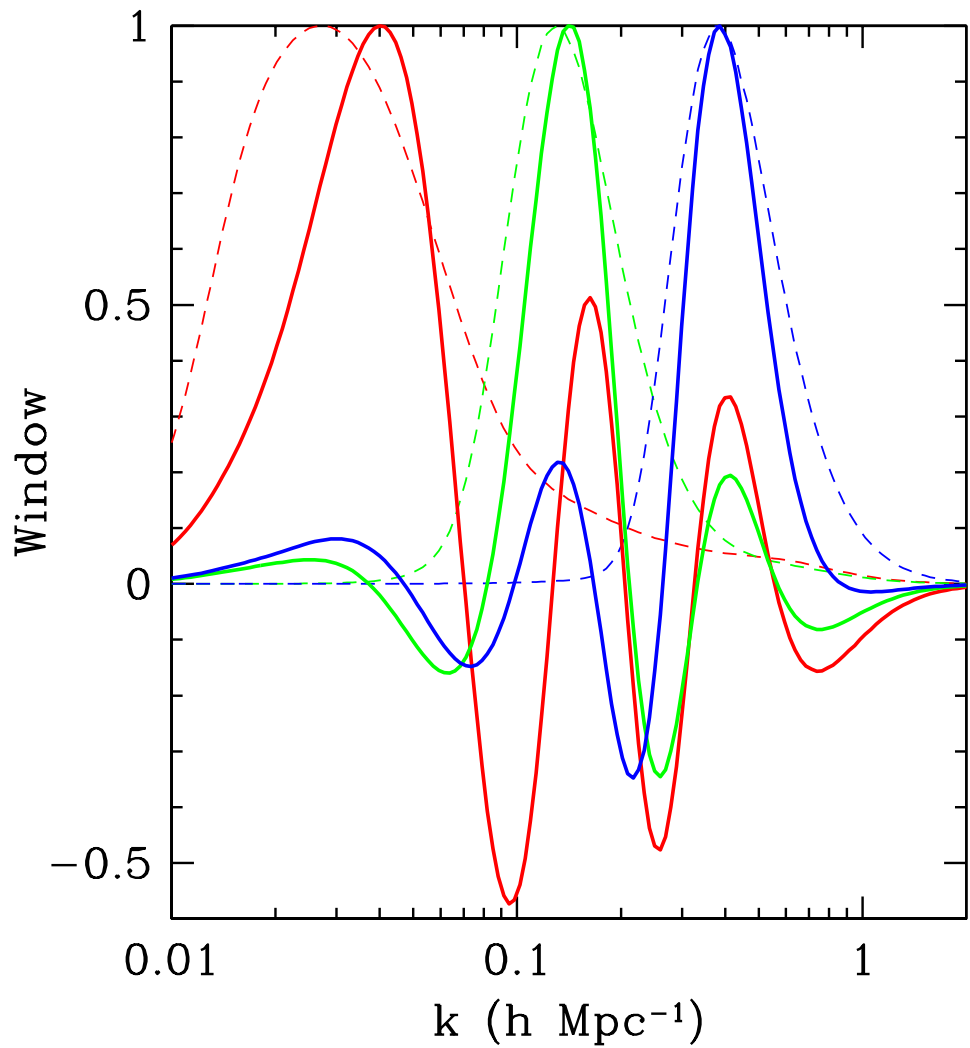}}
\caption{Window functions for 3D power spectrum from the $C_l$'s.
Solid curves are the windows defined via \ec{pest}; dashed are the
windows for the $C_l$'s themselves (equivalent to those in Figure 6 of 
T01). Three different $k$ modes are shown here.}
\label{fig:winmax}
\end{figure}

The estimates of the power spectrum from the $C_l$'s are shown in 
Figure \rf{power_cl}. They do not extend to scales as small as those probed 
by $w(\theta)$ because it is computationally difficult to evaluate 
$C_l$ on small scales. The error bars appear large, but are 
deceptive for two reasons. First, adjacent bins are once again correlated, 
as indicated in Figure \rf{clcov}. The correlations are smaller than those 
we encountered when inverting $w(\theta)$ because, as emphasized by EZ, 
the $C_l$'s are uncorrelated in the all-sky limit, while the covariance 
matrix for $w(\theta)$ is intrinsically complicated. These covariances
persist as we propagate errors through to the power spectrum. 
The second, and more important reason, 
for the large error bars in Figure \rf{power_cl} is that about twice as
many bins were chosen in this case as compared with the $w(\theta)$ inversion,
again to facilitate comparison with the observational window functions.
We will see in \secc{conclusions} that the power 
spectrum from the $C_l$ inversion is almost as constraining as that from 
$w(\theta)$. 

\begin{figure}[htp]
\centerline{\epsfxsize=5truein\epsffile{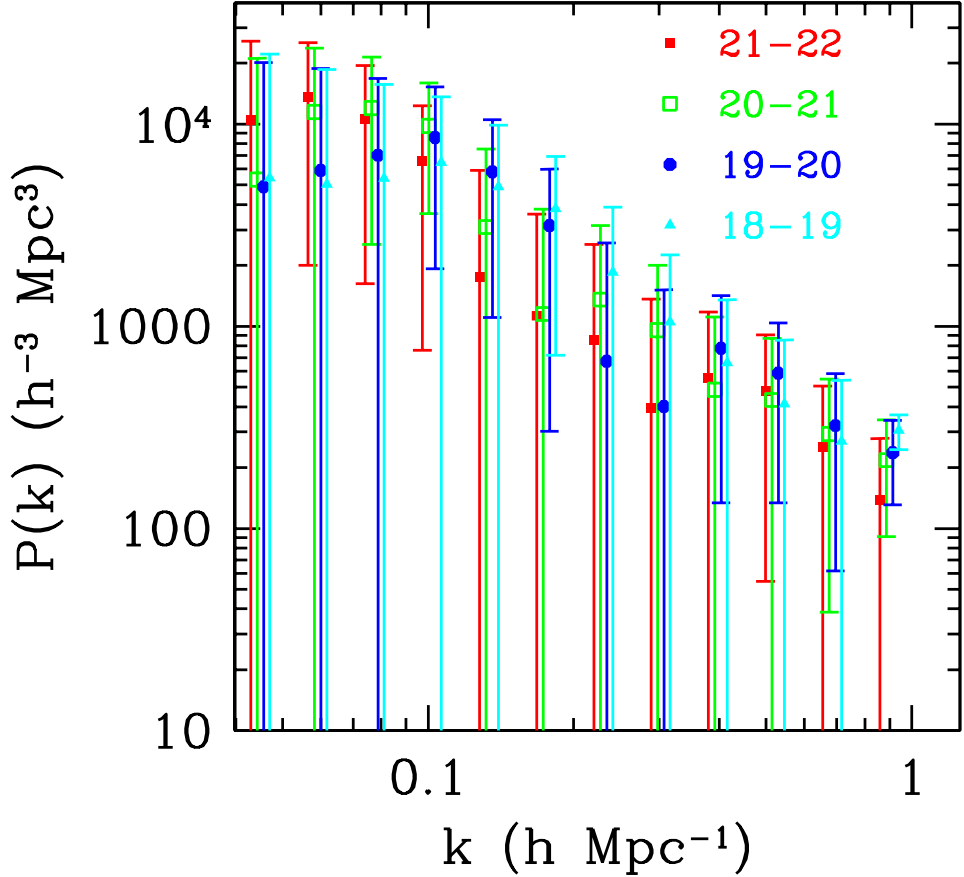}}
\caption{Inferred power spectrum from the $C_l$'s.}
\label{fig:power_cl}
\end{figure}

\begin{figure}[htp]
\centerline{\epsfxsize=7truein\epsffile{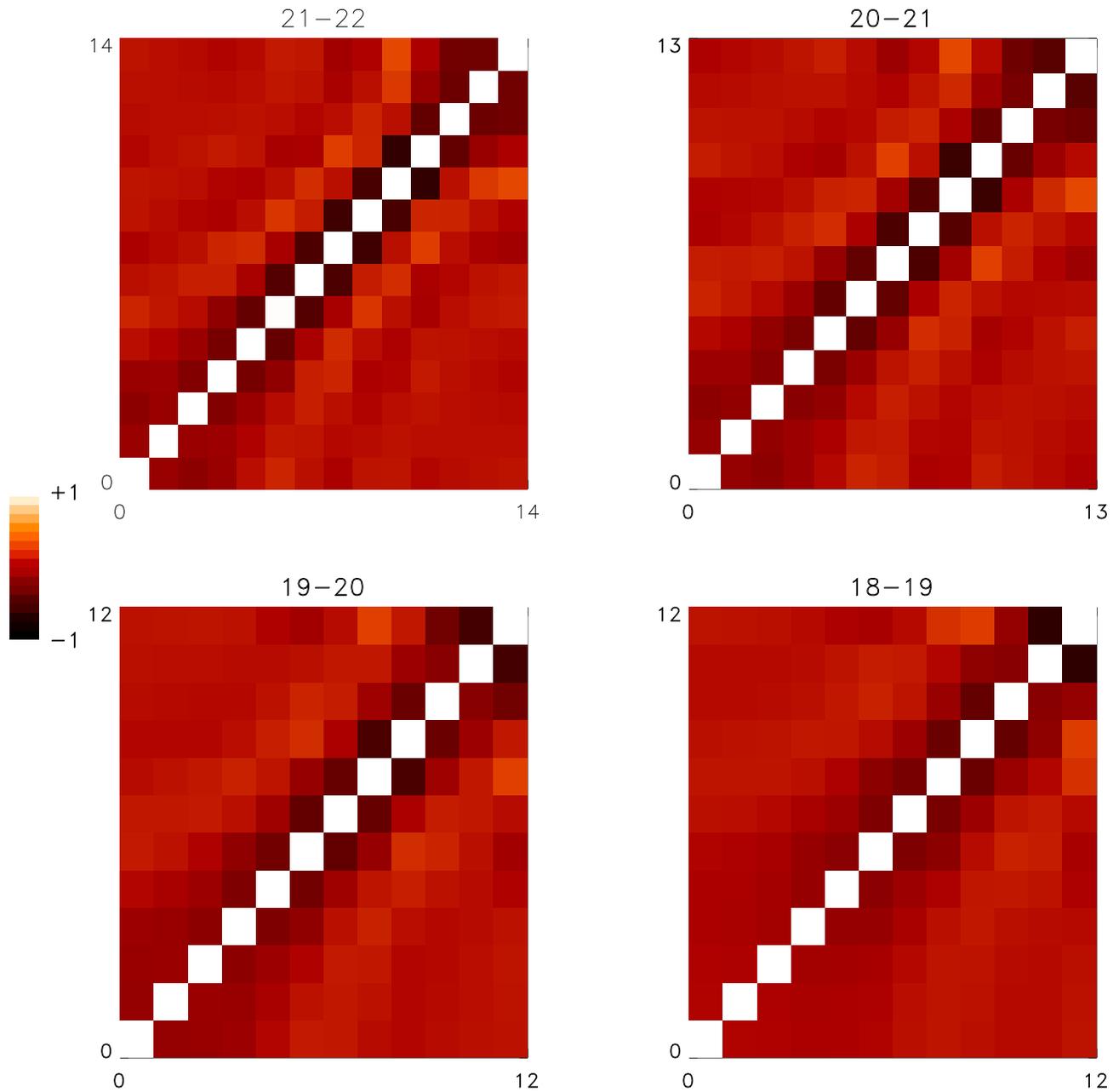}}
\caption{Correlation matrices for $P(k)$ estimated from the $C_l$'s. Again
bottom left corresponds to small $k$; top right to large $k$. These do not
include the $dn/dz$ uncertainty.}
\label{fig:clcov}
\end{figure}

The agreement with the power inferred from $w(\theta)$ is
good, as shown in Figure \rf{clw_comp}. (Here we have used fewer
$k$ bins to allow for easy comparison.) The unmarginalized
errors are shown (primarily to keep the plots from getting too cluttered).
The brightest bin shows how much larger the marginalized errors are in 
each case.

\begin{figure}[htp]
\centerline{\epsfxsize=7truein\epsffile{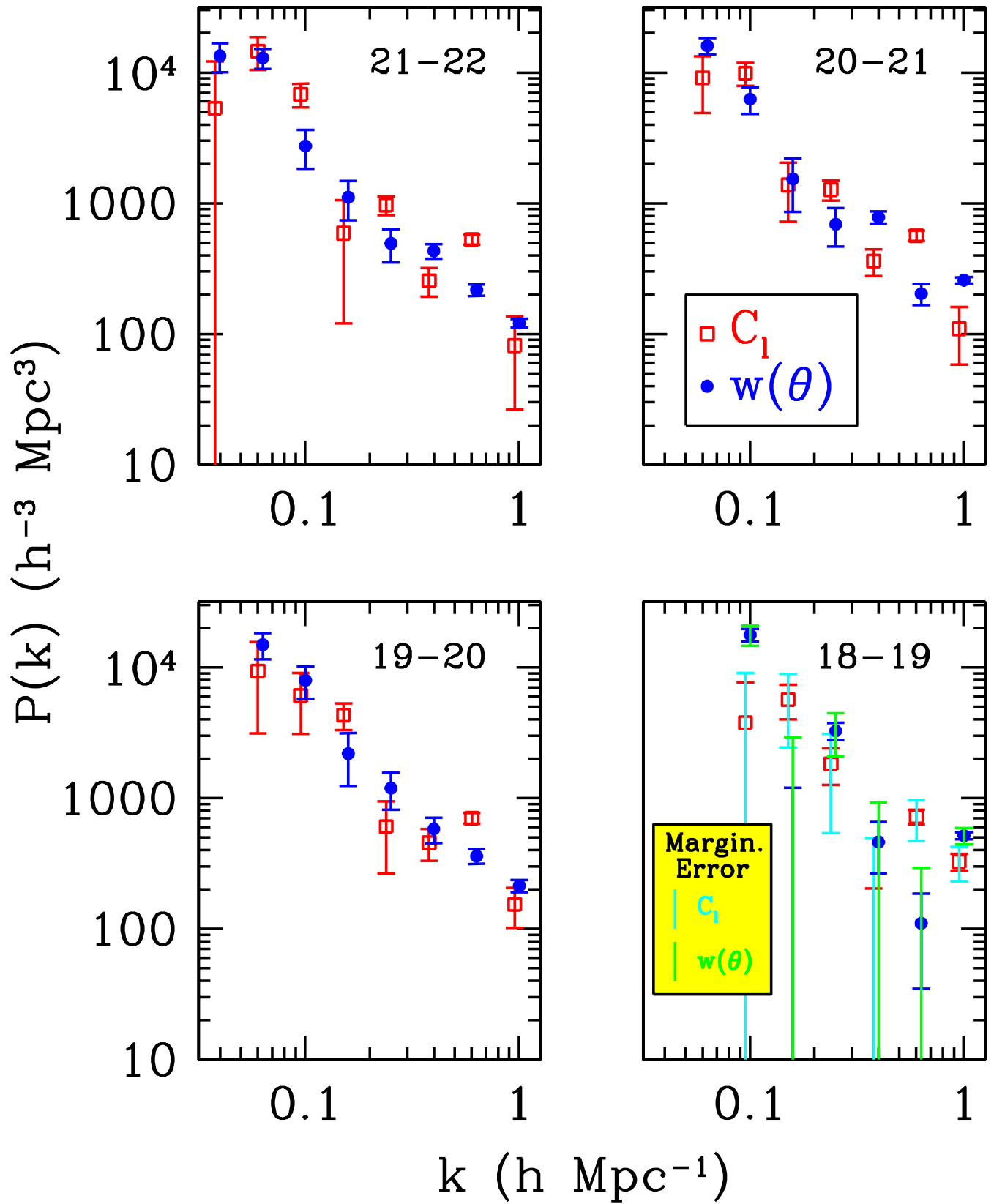}}
\caption{Comparison of the power spectrum from $w(\theta)$ and
from $C_l$ in the four magnitude bins. In this plot
(and this plot only) error bars are unmarginalized errors; non-zero
covariance between adjacent $k$ bins make marginalized errors much larger.
Bottom right panel for brightest bin shows both errors.}
\label{fig:clw_comp}
\end{figure}

The two measures of angular power, $w(\theta)$ and $C_l$, therefore, agree
with each other. This is extremely reassuring given the long path
in each case from raw data to angular power. The small differences are a result
of the many different choices along each route. For example, the quadratic
estimator used by Sc01 and C01 to measure $w(\theta)$ differs from that
used by T01 to measure $C_l$. The latter assumes a prior power spectrum;
 in principle it handles edge effects more accurately. Nonetheless, the 
respective covariance matrices account for the efficiency of the estimator. 
T01 used a Karhunen-Loeve (KL) decomposition
to measure $C_l$, in the process eliminating many low weight modes. 
In a similar vein, the pixelization in T01 was coarser than that of Sc01 
and C01. Both these differences mean that $w(\theta)$ includes
more small scale information since the low weight KL modes are small scale
modes. We do not expect this to lead to any significant differences 
on large scales. 

Perhaps the most difficult part of the computation in 
each case is the covariance matrix.
The final (small) differences between the two sets of measurements are most 
likely due to the different ways of computing these error matrices. 
The $w(\theta)$ covariance matrix comes from the simulations of
Scoccimarro \& Sheth (2001). There are two 
worries with this approach: (i) the underlying model used in the simulations 
may not be correct and (ii) estimating the $N_d(N_d+1)/2$ elements 
from a finite number of simulations (in this case two hundred) is 
dangerous. The second worry is of particular concern because
the off-diagonal elements are non-negligible.
Sc01 however show that at least the diagonal elements
of the covariance matrix agree with the Gaussian approximation on large scales.
The $C_l$ covariance matrix was obtained assuming
the perturbations are Gaussian. The worry here is that non-Gaussianities 
might be creeping in and, for example, inducing non-zero correlations among 
the different bands. T01 show, however, that the kurtosis is consistent
with the Gaussian prediction, suggesting that contamination is unlikely.
While these checks are reassuring, we believe that the small differences that
remain are due to the uncertainties in the covariance matrix. 
Fortunately, these uncertainties will be less significant, 
at least on large scales, for the full SDSS data set which will
go even deeper into the linear, Gaussian regime.

\pagebreak

\section{Conclusion}
\nopagebreak
\label{sec:conclusions}
\nopagebreak

We have analyzed early photometric data from the SDSS and estimated the
3D power spectrum. These estimates are compatible with
previous estimates from other angular surveys, such as APM. There are fewer
galaxies in this early data set than in APM, but the survey goes deeper. One
intriguing result is that the power spectrum inferred from the faintest 
galaxies is lower than that from the brightest. This may be a detection of 
evolution in the power spectrum or an indication of bias or some mixture of these
effects. Since it is most apparent on the smallest 
scales, the excess power in the brightest bins could be a signal of
the onset of nonlinearities.

This paper has focused on one route from the data to the power spectrum
(see Figure \rf{relation}). A companion paper, Szalay et al. (2001),
goes directly from the binned overdensities to a parameterized version of 
the power spectrum. We can compare these two efforts by fitting the
non-parametric power spectrum derived in \S 4 to  
$\Gamma$ and $\sigma_8$ (from Bardeen et al. 1986), which Szalay et al. (2001)
used to parametrize the power spectrum. 
The reduction to just two parameters
also offers yet another way of comparing $w(\theta)$ with $C_l$.

Figure \rf{pkcon} shows the constraints on the parameters 
$\Gamma$ and $\sigma_{8}$ that characterize the shape and amplitude of the
linear theory power spectrum, respectively. For every pair of values of
$\Gamma$, $\sigma_8$, we compute the $\chi^{2}$ between
the model linear theory power spectrum given by Bardeen et al. (1986) and 
the power spectra estimated from $w(\theta)$ and $C_l$ in each of the four 
magnitude bins, using the associated covariance matrices and the window 
functions. The solid circles show the points where $\chi^{2}$ reaches
its minimum value $\chi^{2}_{\rm min}$. The contours are drawn at
intervals in $\Delta\chi^{2} = \chi^{2}-\chi^{2}_{\rm min}$ of
1, 4 and 9. Thus, the projection of these contours on the two axes
correspond to the $68.3\%,\ 95.4\%$ and $99.7\%$ confidence intervals
on the parameters. Note that the contours corresponding to these
confidence intervals on the {\it joint} distribution of the two parameters 
will enclose more area than those shown in the Figure
(see \S15.6 in Press et al. 1992).
Since the shape and amplitude of the power spectrum on 
small scales is altered by non-linear evolution of density fluctuations, we 
restrict the fit of linear theory model power spectra to scales larger
than $k = 0.2\ $h Mpc$^{-1}$, where $\Delta^{2}(k)$ is less than unity
(see Fig. \rf{power_wtheta}).
Red and blue contours show the constraints using the power spectrum 
inverted from $w(\theta)$ and $C_l$, respectively. Green contours show
the constraints derived by fitting predictions of $C_l$ from model linear 
theory power spectra to all the $C_l$ measurements of T01 (using the 
appropriate window functions described there).
In the three bright magnitude bins, we also show the maximum
likelihood estimates of the parameters using the KL technique
of Sz01.

\begin{figure}[htp]
\centerline{\epsfxsize=7truein\epsffile{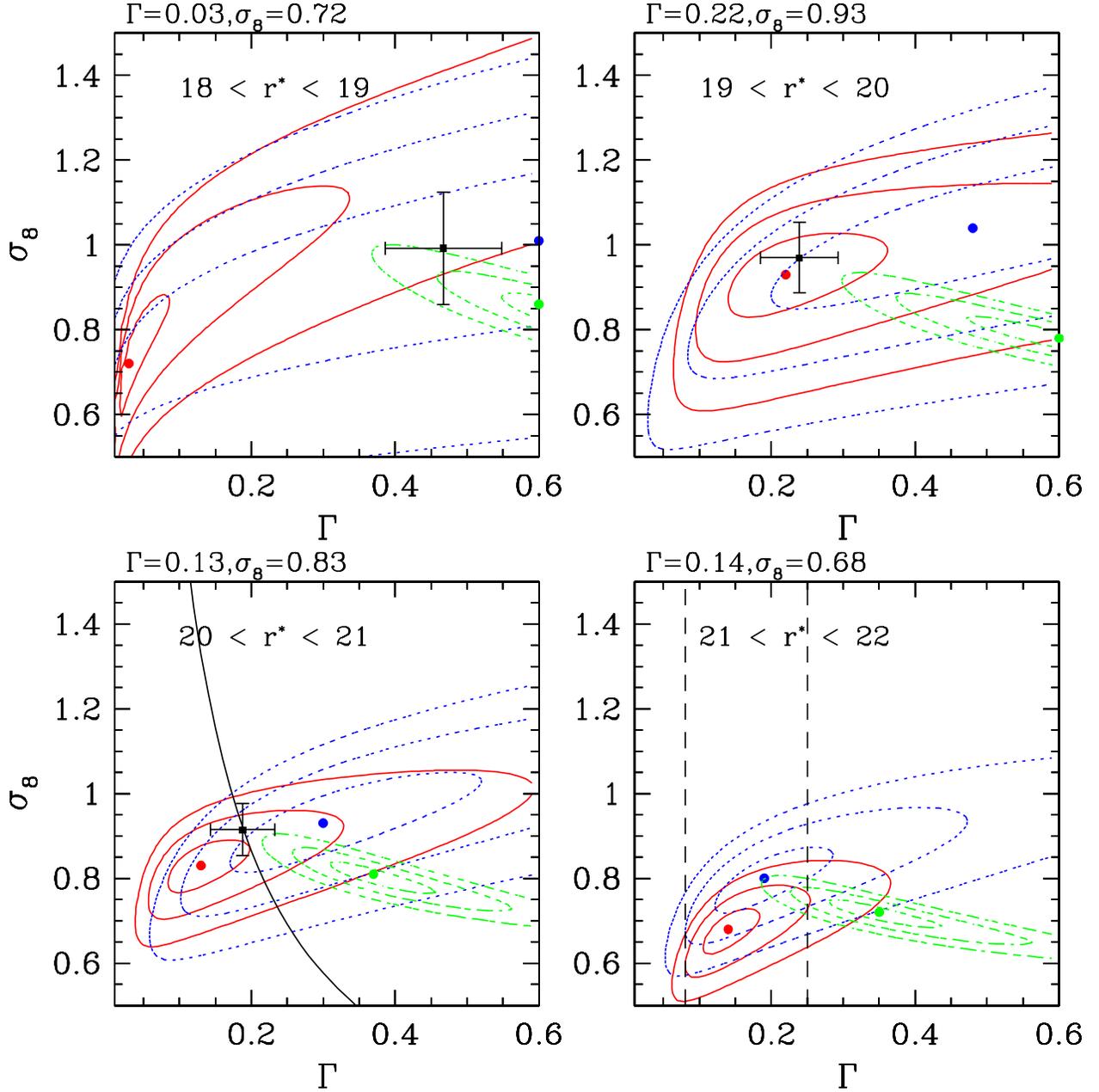}}
\caption{Constraints on $\Gamma,\sigma_8$ using inverted power spectra in 
four magnitude bins. In each panel, red (solid lines) and blue (dotted lines) 
contours show the constraints using the power spectrum inverted from 
$w(\theta)$ and $C_l$
(restricted to scales $k < 0.22$h Mpc$^{-1}$), respectively. Green contours 
(dot-dashed lines) show the constraints using all the $C_l$ measurements 
of T01. The solid square point with errorbars in the three bright magnitude 
bins is the best-fit parameter values and the black curve (solid line) in the 
$20 < r^* < 21$  magnitude bin is the constraint 
$\Gamma\sigma_{8} = 0.173$,
from the KL estimation technique of Sz01.}
\label{fig:pkcon}
\end{figure}

Figure \rf{pkcon} shows that the best-fit values of $\Gamma$ and 
$\sigma_8$ from $w(\theta)$ and $C_l$ are consistent with each other 
in all magnitude bins, with disagreements never exceeding about $1.5\sigma$
(recall that the joint 2D contours are broader than those shown).
In the two intermediate magnitude bins, these best-fit values
are also consistent with the maximum likelihood estimates using the
KL technique from Sz01. It is reassuring that 
three different methods of estimating these power spectrum parameters,
each of which makes different assumptions, yield consistent results.
The contours from the $C_l$ measurements of T01 (green contours)
appear inconsistent with all these three results. 
However, this appearence is illusory, a direct result of
including all the $C_l$ measurements. Many of
these are contaminated by  non-linearities in the 
galaxy power spectrum. As discussed in \S 4.1,
the observational window functions corresponding to $C_l$ are broad,
extending to quite non-linear scales 
(see Fig. 5 of T01).
Indeed, we find that by fitting the linear theory
model power spectrum to the power spectrum inverted from $C_l$ on all 
scales (i.e. not restricting the analysis to $k < 0.2\ $h Mpc$^{-1}$), 
we can exactly reproduce the constraints from directly fitting to the $C_l$ measurements. 

As can be seen from Figure \rf{power_wtheta}, the power spectrum
estimated from the faintest magnitude bin has a smaller amplitude and is
less non-linear than the power spectrum in the brighter bins.
Hence, we expect that the constraints on the shape parameter
$\Gamma$ from this bin is least affected by non-linear evolution.
Therefore, we adopt as our best estimate of $\Gamma$ the value obtained 
from inverting $w(\theta)$ in the
$21 < r' < 22$ magnitude bin. We estimate the error in $\Gamma$ by
marginalizing over $\sigma_8$. Thus, our best estimate of $\Gamma$ is
\be
\Gamma = 0.14^{+0.11}_{-0.06},
\ee
where the error bars are derived by projecting the $\Delta \chi^{2} = 4$
contour onto the $\Gamma$ axis, and therefore correspond to a 
$95\%$ confidence interval.
This value of $\Gamma$ is consistent with the best-fit values
of $\Gamma$ derived by Sz01 in the $20 < r' < 21$
magnitude bin. It is also consistent with the value estimated by
Percival et al. (2001), from the 2dFGRS (see the discussion in
Sz01 for a detailed comparison the 2dFGRS results).

These results on inverting the 3D power spectrum from photometric galaxy
catalogs bode well for the future. The inverted power spectrum using
only the SDSS commissioning data already probes interestingly
large scales. It is reasonable to expect that estimates of the power 
spectrum from all the galaxies expected in the entire SDSS photometric survey,
which will be about fifty to a hundred times larger than this 
early catalog, will be among the most discriminating in cosmology.

{\it Acknowledgements:} 
The Sloan Digital Sky Survey (SDSS) is a joint project of The University of Chicago, Fermilab, the Institute for Advanced Study, the
                         Japan Participation Group, The Johns Hopkins University, the Max-Planck-Institute for Astronomy (MPIA), the
                         Max-Planck-Institute for Astrophysics (MPA), New Mexico State University, Princeton University, the United States Naval
                         Observatory, and the University of Washington. Apache Point Observatory, site of the SDSS telescopes, is operated by the
                         Astrophysical Research Consortium (ARC). Funding for the project has been provided by the Alfred P. Sloan Foundation, the SDSS member institutions, the National
                         Aeronautics and Space Administration, the National Science Foundation, the U.S. Department of Energy, the Japanese
                         Monbukagakusho, and the Max Planck Society. The SDSS Web site is http://www.sdss.org/. 

SD and RS are supported by the DOE and
by NASA grant NAG 5-7092 at Fermilab, and by NSF Grant PHY-0079251.
MT is supported by NSF grant AST00-71213,
NASA grant NAG5-9194 and the University of Pennsylvania Research Foundation.

\end{document}